\documentclass[acmtog]{acmart}

\input epsf
\usepackage{graphicx}
\usepackage{subcaption}  
\usepackage{enumitem}   
\usepackage{xfrac}
\usepackage{epsfig}
\usepackage{booktabs} 
\usepackage{tikz}
\usepackage{color}
\usepackage{hyperref}
\usepackage{float}
\usepackage{setspace}
\usepackage{parskip}
\usepackage{fancyhdr}
\usepackage{pifont}
\newcommand{\cmark}{\ding{51}}%
\newcommand{\xmark}{\ding{55}}%

\usepackage{url}
\usepackage{longtable}
\usepackage{algorithm}
\usepackage{algpseudocode}
\usepackage{tabularx}
\usepackage{gensymb}
\usepackage{acronym}
\usepackage{lscape}
\usepackage{wrapfig}
\graphicspath{ {figs/} }
\usepackage{array}
\usepackage{fancyhdr}

\usepackage[normalem]{ulem}

\newacro{iot}[IoT]{Internet of Things}
\newacro{ai}[AI]{Artificial Intelligence}
\newacro{rfid}[RFID]{Radio Frequency Identification}
\newacro{vr}[VR]{Virtual Reality}
\newacro{ar}[AR]{Augmented Reality}
\newacro{mooc}[MOOC]{Massive Online Open Courses}
\newacro{mcu}[MCU]{Micro Controller Unit}
\newacro{ict}[ICT]{Information and Communication Technology}
\newacro{cps}[CPS]{Cyber Physical Systems}

\AtBeginDocument{%
  \providecommand\BibTeX{{%
    \normalfont B\kern-0.5em{\scshape i\kern-0.25em b}\kern-0.8em\TeX}}}

\setcopyright{acmcopyright}
\copyrightyear{2023}
\acmYear{2023}
\acmDOI{XXXXXXX.XXXXXXX}

\acmJournal{CSUR}
\acmVolume{00}
\acmNumber{00}
\acmArticle{00}
\acmMonth{00}

\begin{document}

\title{Towards Smart Education through the Internet of Things: A Review}

\author{Afzal~Badshah}
\orcid{0000-0002-3444-4609}
\email{afzal.phdcs120@iiu.edu.pk}

\author{Anwar Ghani}
\authornote{Corresponding author of the article}
\orcid{0000-0001-7474-0405}
\email{anwar.ghani@iiu.edu.pk}
\affiliation{%
  \institution{Department of Computer Science, International Islamic University Islamabad}
 \streetaddress{}
 \city{Islamabad}
 \state{}
 \country{Pakistan}}

\author{Ali Daud}
\authornote{Co-corresponding author of the article}
  \affiliation{%
  \institution{Abu Dhabi School of Management}
  \streetaddress{}
  \city{Abu Dhabi}
  \country{United Arab Emirates}}
\affiliation{%
  \institution{Department of Computer Science and Artificial Intelligence, University of Jeddah, Jeddah, Saudi Arabia}
  \streetaddress{}
  \city{Jeddah}
  \country{Saudi Arabia}}
\email{alimsdb@gmail.com}

\author{Ateeqa Jalal}
\affiliation{%
  \institution{Department of Computer Science, University of Science \& Technology Bannu}
  \streetaddress{}
  \city{Bannu}
  \state{KPK}
  \country{Pakistan}
  \postcode{}}
\email{ateeqajalal@gmail.com}

\author{Muhammad Bilal}
\affiliation{%
  \institution{Department of Computer Engineering, Hankuk University of Foreign Studies}
  \streetaddress{}
  \city{Yongin\-si}
  \state{Gyeonggi-do}
  \country{South Korea}}
\email{m.bilal@ieee.org}

\author{Jon Crowcroft}
\affiliation{%
  \institution{Department of Computer Science and Technology, University of Cambridge}
  \city{}
  \country{United Kingdom}}
\email{Jon.Crowcroft@cl.cam.ac.uk}

\renewcommand{\shortauthors}{Afza et al.}

\begin{abstract}

IoT is a fundamental enabling technology for creating smart spaces, which can assist the effective face-to-face and online education systems. The transition to smart education (integrating IoT and AI into the education system) is appealing, which has a concrete impact on learners' engagement, motivation, attendance, and deep learning. Traditional education faces many challenges, including administration, pedagogy, assessment, and classroom supervision. Recent developments in ICT (e.g., IoT, AI and 5G, etc.) have yielded lots of smart solutions for various aspects of life; however, smart solutions are not well integrated into the education system. In particular, the COVID-19 pandemic situation had further emphasized the adoption of new smart solutions in education. This study reviews the related studies and addresses the (i) problems in the traditional education system with possible solutions, (ii) the transition towards smart education, and (iii) research challenges in the transition to smart education (i.e, computational and social resistance). Considering these studies, smart solutions (e.g., smart pedagogy, smart assessment, smart classroom, smart administration, etc.) are introduced to the problems of the traditional system. This exploratory study opens new trends for scholars and the market to integrate ICT, IoT, and AI into smart education.
\end{abstract}

\keywords{\ac{iot}, Education System, Smart Education, E-Learning, Flipped Classrooms}

\maketitle

\section{Introduction}
\label{intro}

Smart education (also referred to as education 4.0) is a teaching, learning, and managing paradigm, where smart technologies (e.g, \ac{iot}, \ac{ai}, and 5G) are applied to make it more effective and attractive \cite{SEI193,SEF114, SEF113}. Smart education provides a digital environment to facilitate learners, parents, teachers, and administrators to enhance learner engagement and motivation \cite{SEI156}. For smart education, different terminologies are used, such as smart university, smart learning, smart classroom, smart learning environment, etc \cite{SEI171,SEI196}.  The word "smart" refers to intelligence in decisions, penalization (in learning and teaching), transparency, and adaptivity in smart education. With the recent development in smart technologies, \emph{educational institutions} need to be transformed into smart institutions rather than continue with the traditional methods \cite{SEF113,SEI197}.  Table \ref{tab:steducation}  differentiates smart and traditional education.

 \begin{table*}[ht]
\centering
\caption{Analysis of smart and traditional education}
 \label{tab:steducation} 
\newcolumntype{b}{X}
\newcolumntype{s}{>{\hsize=.6\hsize}X}
\newcolumntype{c}{>{\hsize=.2\hsize}X}
\setlength{\extrarowheight}{2pt}%
\begin{tabularx}{\textwidth}{@{}s s@{}} 
\toprule
\textbf{Smart Education}  &  \textbf{Traditional Education}  \\
\midrule

Flexibility is an influential advantage of smart education.  It allows students to learn in their own space.  Teachers can listen repeatedly \cite{SEI157}. &	Compared with smart education, traditional education is not so flexible. Students must go to classes \cite{SEI161}.  \\

Smart education is not location and time-dependent. Students can watch lectures anywhere and anytime  \cite{SEI158}.	& Concerning smart education, traditional education is location and time-dependent and has very firm policies. \\

Operational cost is a big issue for institutions and learners. Smart education minimizes this cost in many ways; for example, paperless work minimizes the expenditure of travelling, examination and administration, etc.	& Traditional education needs more operational costs than smart education in terms of paperwork, travelling expenditure, examinations, administration, etc. \\

Smart education allows teachers to teach worldwide; therefore, a large number of courses can be offered\cite{SEI159}. &	Traditional education does not follow this case. It needs the physical presence of teachers. Therefore, it cannot offer every type of course.   \\

There are a number of channels (Zoom \cite{SEI141}, Google meet  \cite{SEI142}, Skype  \cite{SEI143} etc), which are used for communication and collaboration.  	&   In the case of traditional education, physical presence is needed, and no other options are available. Especially in the meeting, a massive budget is consumed in daily and travelling allowances.   \\

The major lack of smart education is social interaction. Learners stay home or remain busy with gadgets, etc., which badly affects their social and communication skills  \cite{SEI160}.  &	One of the major benefits of traditional education is social learning. Students get together and participate in face-to-face discussions, which increases their social skills \cite{SEI161}.  \\

The other issue with smart education is that it does not have extracurricular activities. &	Instead of smart education, traditional education provides extracurricular activities. This prepares the learners for further study \cite{SEI162}.\\

Although smart education minimizes operational costs, it needs a massive capital cost to install a smart system. &	The overall cost of traditional education remains on the high side.  \\

\bottomrule    
\end{tabularx}
\end{table*}

It is estimated that around 75 billion smart devices will be connected to the internet by 2025 \cite{SEI28}. This is a revolutionary hypothesis about the future. Moreover, the launch of the 5G mobile network introduces real-time communication \cite{SEI44}. These large number of devices and super-fast connectivity show that everyone will have a smartphone to communicate in real-time in the near future. These smartphones and smart devices can be used for smart teaching and learning \cite{SEI90}. The Internet of Educational Things (IoET) is proposed for the devices used in smart education, which is missed in the literature \cite{SEI138}.



\begin{figure*}[ht]
 \center
	\includegraphics[width = \textwidth]{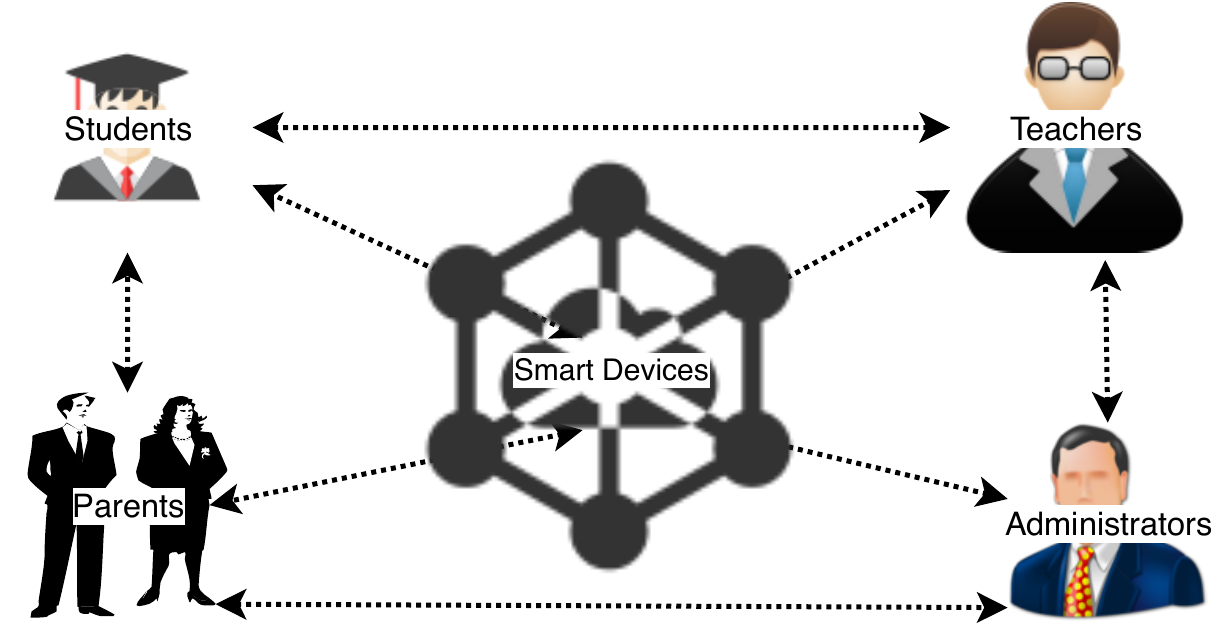}
	\caption{Possible structure of \ac{iot} devices in smart educational institutions}
	\label{fig:smartaaministration}
\end{figure*}

Three essential pillars of the educational management system are;  i) administrators, ii) teachers, and iii) parents. All these stakeholders’ active involvement and productive decisions are crucial for learner development. 
Smart devices help administrators, teachers, and parents to involve learners in learning activities, and to monitor the teaching and learning process in real-time \cite{SEI29,SEI203}.  Fig.  \ref{fig:smartaaministration} shows the possible structure of smart education. In this structure, smart devices are used to connect students, teachers, parents, and administrators.

In the vast majority of institutions, the Standard Operating Procedures (SOPs) are not followed due to a lack of resources, which declines the quality of education.  Furthermore, learners are not creatively engaged in producing a creative generation. \ac{iot} may play a big role in these issues as it has been playing smart tasks in various fields. Unfortunately, it has not been properly integrated into the educational system. Furthermore,  smart education is still not properly included in the \ac{iot} platform ( e.g,  smart cities, smart agriculture, and smart health care, etc); therefore,  there must be a term \emph{"smart education"}. As per our research, this article is the first step toward exploring the possible uses of \ac{iot} in the education system.  This explores all possible uses of \ac{iot} devices within the smart educational institution. \ac{iot} in education enhances the teacher’s and learners' response, performance, and behaviour. Flourishing creativity is an essential goal of education; however,  the current education system (usually in underdeveloped countries) is based on reading, watching, and listening, through which only 2 to 3 percent of information is retained in memory and the rest is forgotten \cite{SEI116}. Learning retention can be increased by up to 90\% by engaging children in practical work, discussion, and teaching. Edgar Dale's research concluded that students retain 10\% from reading, 20\% from listening to a teacher, 30\% from watching, 50\% from watching and listening (at a time), 70\% from participating in discussions and doing practical work and others, and up to 90\% of information remains part of the memory from teaching to others \cite{SEI116, SEI151}. This is a brilliant idea; however, today's teaching methodology misses it. Smart devices can play a role in deploying this learning and understanding the idea \cite{SEF115}. 

Unfortunately, in most regions of the world \cite{SEI132}, the traditional way of teaching and learning is followed, which prepares learners for \emph{a manufacturing-based economy}. The teaching process is not the same as years ago. Everything has changed, including curriculum, pedagogy, assessment,  and teachers and learners. The digital world is affecting all these terms very sharply. Therefore, it is the need of the day to find a smart system that works well for learners, parents, teachers, and administrators. It will empower parents, teachers, and administrators to develop a generation having high values and global skills. Smart institutions not only save the cost but also make the classroom transparent and improve the quality of education \cite{SEI32, SEI152}.

Therefore, the objectives of this research study are;

\begin{itemize} 

\item To explore the challenges of the traditional education system and propose possible solutions in light of the recent \ac{ict} trends (e.g, \ac{iot}, \ac{ai}, and 5G).
\item To investigate new directions to integrate the recent \ac{ict}  trends  with the educational system.
\item To study the resistance and challenges (e.g., computational and social) to the implementation of smart education. 

\end{itemize}

 Smart education got very little attention; therefore, only a few review studies can be found in the literature. However, these studies have poor coverage, only looked at the related studies, and lack depth.  \citet{SEI78} and \citet{SEI79} proposed smart solutions; however, they missed the smart education structure, smart solutions classification, and challenges with the smart system. 
Unlike the existing studies, this survey is covers all aspects of smart education system in depth.  Along with a review of related studies, it addresses the challenges of the traditional education system, proposes possible solutions,  and discusses potential challenges to smart education (e.g., social and IT challenges).


The general contributions of this article are listed as follows:

\begin{itemize} 
\item \emph{Reviewed} the important investigations, projects, and frameworks of \ac{iot}, \ac{ai}, and 5G in the context of smart education by explaining their motivations, challenges, and opportunities. 

\item Explored the \emph{challenges} in traditional educational system in terms of teaching, learning, and management; and \emph{possible smart solutions} to the challenges, e.g., smart classes, smart pedagogy, smart assessment, smart monitoring, and portfolio etc.

\item Proposed \emph{a Smart Human Resource Model} for educational institutions enabling integration of smart devices to manage institutions smartly and \emph{possible use of \ac{iot}, \ac{ai}, 5G technology, and smart devices} in educational institutions to make their activities smart. 

\item Explored the possible \emph{challenges and resistance} from employees, parents, society, and customers along with the possible solutions to handle these challenges and hurdles. Furthermore, it discusses the possible future directions of smart education. 

\item Provided Comparative analysis of the proposed study with related reviews and surveys to demonstrate the supremacy of this study. 
\end{itemize}

To the best of our knowledge, this is the first unique study, reviewing the existing work and proposing the structure of smart education.

The rest of the article is structured as follows:
 Section \ref{ct} explains the concept and terminologies used in this study.
 Section \ref{bg} presents the \ac{iot}, its applications, and explores the \ac{iot} platform which includes the \ac{iot} hardware, connectivity, and software. It further covers the human resources involved in smart education;
 Section \ref{rm} covers the research methodology used in this paper to collect the related literature;  
 Section \ref{literature} classifies the related studies;  
 Section \ref{uses} investigates the  use of smart devices in educational institutions; 
 Section \ref{app} explores the possible use  of smart devices and \ac{iot} in educational institutions; 
 Section \ref{challanges} discuss the possible challenges and resistance to smart educational institutions; 
  Section \ref{ca} compare this study with related surveys and reviews to confirm the supremacy of this work; 
 and finally,  Section \ref{conclusion} concludes the study along with future directions.

 \begin{table*}[ht]
\centering
\caption{List of Abbreviations and Notations}
 \label{tab:abb} 
 \newcolumntype{b}{X}
\newcolumntype{s}{>{\hsize=.4\hsize}X}
\newcolumntype{c}{>{\hsize=.2\hsize}X}
\setlength{\extrarowheight}{2pt}%
\begin{tabularx}{\textwidth}{@{}c s c s@{}} 
\toprule
\textbf{Abbreviations}  &  \textbf{Description} & \textbf{Abbreviations}  &  \textbf{Description}  \\
\midrule

ICT  & Information and Communication Technology	  &	LMS  & Learning management System  \\

CC  & Cloud Computing	  &	MOOC  & Massive Online Open Courses  \\

5G   &	5th Generation Technology (wireless)  &	 MCU &  Micro Controller Unit\\

AI   &	Artificial Intelligence  &	GSM  & Global System for Mobile  \\

 VR     &  Virtual Reality  &	SIM	  & Single Identity Module \\
 
 AR  & Augmented Reality	  &	NB-IoT  & NarrowBand-Internet of Things   \\
 
 IOT    & Internet of Things	  & WAN	  & Wide Area Network   \\   
 
RFID  & Radio Frequency Identification	  & PDAs	  & Personal Digital Assistance \\

GPS & Global Positioning System	  &	SOPs  & Standard Operating Procedures \\
IoE & Internet of Every things  &	SSF  & Smart Security Framework \\
IoLT &	 Internet of Living Things   &	BYOD  &  Bring Your Own Device\\
 IoET &Internet of Educational Things & HDV & High Definition Video \\
\bottomrule    
\end{tabularx}
\end{table*}

\section{Concepts and terminologies}
\label{ct}

This is a multidisciplinary study covering education and \ac{ict}. Therefore, to make it easy for both disciplines, this section is divided into two subsections;  (i) Educational terminology and (ii) Smart system terminology. These sections cover the basic concepts and terminologies used throughout this article. Furthermore, Table \ref{tab:abb} shows the important abbreviations and notations used in this paper.  
 
 \subsection{Educational terminologies}
 
This subsection covers the important terminologies used in the education sector. 

\subsubsection{Pedagogy}

Pedagogy deals with teaching methodology skills. The term pedagogy is broad and includes not only basic terms related to classrooms (e.g, teaching, learning, and assessment skills), but also includes observational skills, feedback, psychology, etc \cite{SEI29}.

\subsubsection{Lesson plan}

Before the start of the class, a written plan is drawn up for a specific topic to use the lesson time effectively. This helps the teachers to manage the classroom effectively and productively. The lesson plan is a blueprint that guides the teacher to lead the class \cite{SEI29}.

\subsubsection{Assessment}

Assessment is the method used to evaluate students learning during or after the class or term. Assessment results are used to take final decisions related to the individual student, teacher, and long-term decisions relating to the institution \cite{SEI30}.

\subsubsection{Portfolio}

A portfolio is a collection of students' and teachers' project records, results, and other activities.  A portfolio helps in future decisions and can be used to train the AI algorithm~\cite{SEI22}.

\subsubsection{Engagement}

In educational institutions, the word engagement means the productive involvement of the learner in the classroom. Productive involvement means that students learn in the classroom instead of wasting time on other activities \cite{SEI30}.

\subsubsection{Flipped classroom}

The flipped classroom is the reverse of the traditional classroom, in which learners watch recorded videos at home and do related activities at school under teacher supervision \cite{SEI88}. 

\subsubsection{Personalized learning}

Personalized learning gives learners the choice of what they want to learn. This positively motivates them to learn and improves their involvement in productive learning \cite{SEI93,SEI195}.

\subsubsection{Activities}

Activities involve learners productively in learning. This involves them cognitively and physically in a task for the productive use of time. Research shows the major positive effects of activity-based learning~\cite{SEI153}. 

\subsubsection{Question bank}

Teachers prepare new questions every time when they take the exam. To manage it smartly, all questions of the topics are stored in a database and retrieved for question making. In a smart assessment, the application automatically populates the questions for assessment~\cite{SEI106}. 
\subsection{Smart system terminologies}

This section covers the important terminologies used in the ICT sector. 

\subsubsection{Smart}

The word smart means to intelligently handle some situations using smart technologies. The word "smart education" refers to handling educational institutions using the \ac{iot}, \ac{ai}, and other related smart technologies \cite{SEI154}.

\subsubsection{Internet of Educational Things (IoET)}
IoET refers to all those devices used in smart education that is connected to the Internet. These devices include sensors, cameras, RFID tags, and other connected devices that can be used to collect data, monitor student activity, and provide feedback to teachers and administrators.  \cite{SEI138}. 

\subsubsection{Internet of Things}

\ac{iot} is the web of connected devices using the wireless network and internet connections. It uses embedded technology and code to collect data or take automatic decisions \cite{SEI106}. As the name suggests, IoT is a large cloud of smart things. These devices and sensors are connected to each other for communication. Now, these devices are integrated with AI to make smart decisions \cite{SEI118}.

\subsubsection{MOOC}

\ac{mooc} (e.g, Coursera \cite{SEI144} and edX \cite{SEI145}) provide online courses facility. Now, with advanced technologies, universities offer \ac{mooc} courses and students receive certificates at home. This is the future trend of teaching, learning, and certification \cite{SEI22}.

\subsubsection{Learning Management System (LMS)}

LMS is an online application used to interact with educational institutions for classes, assignments, quizzes, assessments, etc \cite{SEI22}.

\subsubsection{Virtual classroom}

In virtual classrooms, learners remotely attend online classes on the Internet. With the emergence of 5G (having very minimum delay), virtual classrooms are becoming very popular \cite{SEI22}. 

\subsubsection{Collaboration}

5G has made remote collaboration very easy and smart. Like virtual classrooms, they can use various applications (Zoom \cite{SEI141}, Google meet  \cite{SEI142}, Skype  \cite{SEI143} etc) to collaborate remotely on their subjects \cite{SEI155}.

\subsubsection{Artificial Intelligence}

The \ac{ai} enables smart machines to sense the environment and take decisions as per input \cite{SEI117}. AI is being utilized to comprehend and uncover the best methods of teaching using various forms of data, such as portfolios, classroom or tool-based recorded lectures, and so on~\cite{SEI186, SEI187}. 

\subsubsection{Virtual Reality (VR)}

VR provides a three-dimensional view and can be applied in entertainment and education. VR is now widely used for virtual meetings, training, collaboration, and lectures \cite{SEI119}. VR is making learning easier, faster, and more engaging.

\subsubsection{Augmented Reality (AR)}

AR can be defined as a system that fulfils three basic functionalities: a combination of real and virtual worlds, real-time interaction, and accurate 3D capture of virtual and real objects \cite{SEI119}. AR may be utilized to enhance the learning process in both online and offline teaching techniques by allowing the linking of imagination with the real world through virtual means. 

\subsubsection{Fifth generation (5G)}

The fifth generation of communication technology provides high-speed network connectivity, which enables real-time communication. Real-time communication is essential for the adoption of AR and VR technologies in interactive learning and training sessions.

\subsubsection{MCU}

MCU stands for micro-controller unit which works as a hub for smart sensors. The sensors are connected with the MCU, and it further communicates this data with fog and cloud computing \cite{SEI147}.

\subsubsection{Fog computing}

Instead of processing data on the cloud to overload the network and cloud computing, the data is processed on edge server. Processing or storing data on edge devices is called fog computing \cite{SEI121}. 

\subsubsection{Cloud computing}

Cloud computing provides desktop computer resources online on the network. These services are categorized as Infrastructure as a Service (IaaS), Software as a Service (SaaS), and Platform as a Service (PaaS) \cite{SEI122,SEI180}.


\section{Background}
\label{bg}

The web of \ac{iot} is getting wider day by day. With the increase in internet speed and coverage, the growth of smart devices is also increasing. Giant tech companies and governments are working together to enhance internet coverage in remote areas using various technologies. Google is working to provide the Internet to far-flung areas using \emph{balloon stations} \cite{SEI31}, CISCO has a big share in communication devices. Furthermore, SpaceX is launching satellites to directly provide the Internet to remote areas from lower orbit satellites. The difference between Starlink (SpaceX) and other geostationary satellites is that these satellites are located 35,000 km from Earth, which has a higher delay, however, the Starlink satellites orbit is about 550 km and provide a fast Internet connection. Although this technology is still in its infancy, however, marketing internet connections has already started \cite{SEI189}. With this speedy growth, it is expected that more than 75 billion devices will be connected to the internet by 2025 \cite{SEI28}. This figure is about nine times greater than the world population. This section covers the discussion on various smart education-enabling technologies.

\subsection {Internet of Things (IoT)}
\label{iot}



The \ac{iot} allows us to create smart systems by connecting smart gadgets. Normally, any institution or building is smartly managed by hundreds of devices. Every electric appliance in a smart home, such as a refrigerator, lamp, fan, security camera, kitchen appliance, and other equipment, has its own sensors \cite{SEI53}.

\ac{iot} improves device connectivity and fosters communication, teamwork, and experience sharing. To enable these devices to comprehend their surroundings and make critical judgements, artificial intelligence and AIOT are being combined \cite{SEI204}.

 \begin{figure}[ht]
 \center
	\includegraphics[scale = .4]{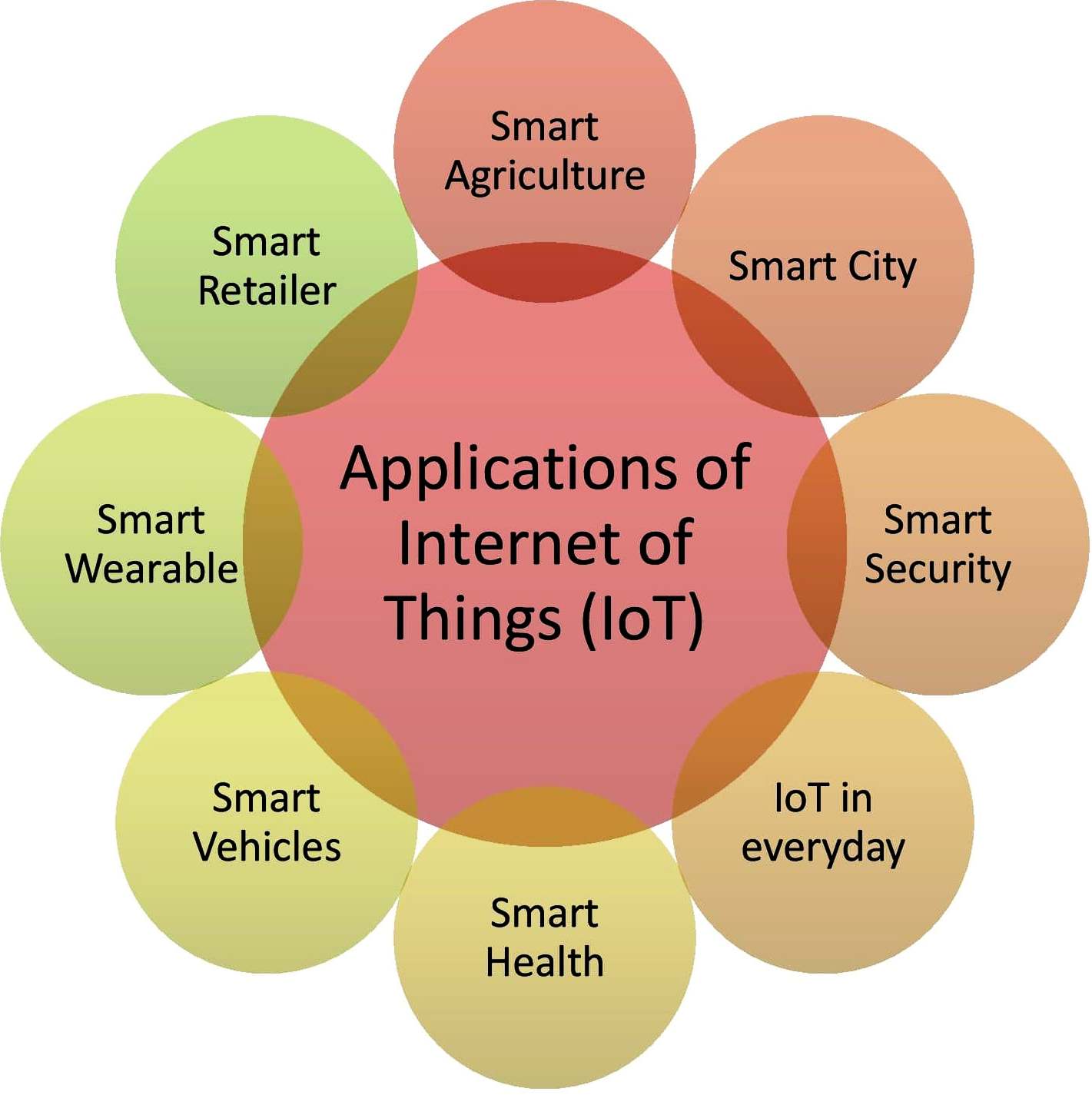}
	\caption{Applications of \ac{iot} in different fields}
	\label{fig:application}
\end{figure}

\subsection{Applications of IoT }
\label{application}

\ac{iot} is  used in every field of life, as shown in Figure  \ref{fig:application}. This new wave of smart technology goes beyond smartphones, tablets, or sensors. Today, \ac{ai}  is integrated into tiny devices, which empowers them to make automated decisions. This massive flood of smart devices around the world is opening several new doors and introducing new paradigms \cite{SEI16}. 

One of the powerful applications of \ac{iot} devices in \emph{everyday life} is to read the surrounding environment and respond according to the sensed data. For example, handling \emph{agriculture} traditionally is a waste of time and money. \ac{iot} manages these issues with smart systems. The smart watering system saves water, and if water is needed for any field or crop, the pump automatically starts watering it. There is no human interference, and this is done automatically.  In the greenhouse, the sensors read the environment, and it is automatically controlled \cite{SEI41}. 

\emph{Smart industries} are the major part of \ac{iot} application. Instead of creating a product manually, robotic hands commercially produce the product automatically. This drastically increases the speed, accuracy, and reliability of production.  \ac{iot} optimizes the time and cost;  increases the quality and production of products \cite{SEI40,SEI70}.

IoT helps in medical examination and analysis of the results and empowers the \emph{health care} system to improve medical reports. This reduces the time and cost invested in the health sector.  Smart devices are attached to critical patients.  If the patients’ condition goes down, the sensors generate a notification which is communicated to the physician and relatives \cite{SEI43,SEI179,SEI142}. 

\ac{iot} applications are used to control city traffic, parking,  waste management,  security, and safety systems, and many more. The environment is read with \ac{iot} sensors to give a good environment to the citizens \cite{SEI38}.   

IoT provides a good solution for \emph{security}. Sensors and security cameras detect the insecure situation and alert concerned officials.

\begin{table*}[ht]
\centering
\caption{ Comparative analysis of Micro Control Units (MCUs) used in market}
 \label{tab:mcusummary} 
 \newcolumntype{b}{X}
\newcolumntype{s}{>{\hsize=0.4\hsize}X}
\setlength{\extrarowheight}{2pt}%
\begin{tabularx}{\textwidth}{s b s s s s s s} 

\toprule
\textbf{Board} &  \textbf{Summary} & \textbf{Processors} &  \textbf{Processor Speed} & \textbf{Size (mm)} &  \textbf{USB Port} &  \textbf{Memory} &  \textbf{Price}\\
\midrule

 Arduino Uno \cite{SEI34}  &  Current "official" Arduino USB board, driverless USB-to-serial, auto power switching  &  ATmega328   &  16 MHz  & 73 x 53   & USB 2.0  & 1 GB & \$ 29.99  \\
 
Raspberry Pi \cite{SEI35}   &  Single board Linux computer with video processing and GPIO ports  &  ARM 1176JZF-S   & 700 MHz  & 85x 56  & USB 2.0 & 64 MB &  \$ 39.99 \\

CloudBit  \cite{SEI36}  & Customized Arch Linux ARM, No video processing  & MX233  & 454 MHz  & 55 x 19  & No & 64 MB & \$ 59.95 \\

UDOO  \cite{SEI37}  &  UDOObuntu, Android, XMBC, Yocto, Arch Linux, OMV  & ATmega 256RFR2  & 1 GHz  &  110 x 85  & USB OTG USB 2.0  & 1 GB  &  \$135 \\
\bottomrule    
\end{tabularx}
\end{table*}

\subsection{IoT platform  }
\label{platform}

\ac{iot} platform is categorized into three different layers. These are \ac{iot} hardware, \ac{iot} connectivity, and \ac{iot} software.  \ac{iot} hardware, connectivity and software create a fleet of \ac{iot} devices (\ac{iot} Cloud). 

\subsubsection{Physical layer}

\ac{iot} hardware consists of smart boards, sensors, actuators, and other smart devices. The core component of \ac{iot} is \ac{mcu}, which gathers data from other associated devices and interprets it to make various logical decisions. The sensors may be installed in the \ac{mcu} or they work separately. Some health care and security-related gadgets are also worn \cite{SEI56, SEI208}.  
There are various companies that manufacture and do research on \ac{iot}  hardware. Table \ref{tab:mcusummary} shows the detailed comparative analysis of smart system circuit boards.

\emph{Arduino} is a single-chip open sources project. Its hardware and software are free to be used or modified for different projects. Arduino has various circuit boards with different capabilities, for example, it comes with onboard Wi-Fi, Ethernet, and USB connections. It uses Linux as the operating system. Arduino's simple programming functions are used to program it for different tasks \cite{SEI11}.

\emph{Raspberry Pi} is a single-board computer chip. It is an open-source project and can be freely modified for any project.  Raspberry Pi provides a range of single-board computers which are used according to the requirements. Raspberry PI 4 provides a choice for RAM up to 4 GB; a powerful microprocessor is installed. Apart from that, it provides a USB port, Wi-Fi port, and other micro HDMI ports  \cite{SEI12}.

\emph{Cloudbit-little bit} uses the littleBits circuit, such as the button placed on the door to be opened. When the littleBits button is pressed, it generates a message which is sent to the required destination. In this, small other modules work, which can be connected according to the need. All these modules are magnetically attached to each other. This facility makes it more attractive  \cite{SEI13}. 

\emph{Particle} works on all three main pillars of the \ac{iot}. They work on devices that have hardware, software, and interconnections. They provide cellular and Wi-Fi connections in their particle boards. This is the main benefit of the particle, which other companies do not provide. They also handle security, scalability, and reliability  \cite{SEI14}.

\emph{UDOO} was developed in 2013. This is the next step to develop a power full \ac{iot} connectivity. This provides different types of boards such as DUAL/QUAD, NEO, X86, and BLU. UDOO is well in a web-based platform for monitoring and organizing  \cite{SEI15}.

\emph{Sensors and actuators} are smart devices that read the physical changes in the environment. These readings are sent to the  \ac{mcu}. \ac{iot} has a variety of sensors such as motion sensors, temperature sensors, frequency identification, sound detectors, water detectors, smoke detectors, and many more  \cite{SEI16, SEI173}.

\subsubsection{Communication layer}

Connectivity is the foundation of \ac{iot}. It is a key decision to select appropriate devices for connectivity. This may increase the cost or reduce the performance in case of a wrong selection. Connectivity device selection is made on behalf of range, data throughput, energy efficiency, and device cost. Table \ref{tab:connectummary} shows the detailed comparative analysis of connectivity technologies (wireless).

\ac{mcu} uses \emph{GSM} circuits and cellular phone SIMS to connect the object to the internet, which provides long-distance control to the objects.  \emph{Wi-Fi}  (802.11), is a wireless protocol for short-range connections having a range of about 100 m and performs well in short ranges.   \emph{Ethernet} connects the devices with a physical wire,  usually used to directly connect the device. Unique IPs may be used if required.   \emph{Bluetooth} is a short-range connectivity technology,  now rarely used after the emergence of Wi-Fi. \emph{ZigBee} is a low-power WAN, used for smart system communication.  The use of ZigBee has several advantages over other networks because it is simpler and cheaper. The transfer distance of ZigBee is more than 100 m.  \emph{Narrowband IoT  (NB-IoT)}  is a pioneering technology in \ac{iot} communication. It covers more distance and consumes less power compared to Wi-Fi. 

\subsubsection{Software layer}

A range of languages are used to develop \ac{iot} applications. For circuit board programming, mostly Java and Python are used. They have built-in functions, which are employed with these languages. Linux or Windows 10 is used as an operating system in \ac{mcu}. Apart from these, other OSs may also be used in different \ac{mcu}s.

\subsection{Human resources}
\label{hr}

This section discusses the stockholders involved in the teaching-learning process.

\emph{Learners} are the core entity for which all this is being done. Today children are the next leaders. They prepare the next society. If we train them as good humans today, surely, they will form a sound human society of tomorrow. 

 Any country's education system is as important as its \emph{teachers}. Teachers are the core of the learning process.  Empowering teachers with smart technologies can improve the education system manyfold.  By integrating smart technologies into the education system, teachers can adopt smart technology to innovate their teaching methods and make their classes more interesting, engaging, and effective \cite{SEI212}.  
 
\emph{Administrators} take decisions on behalf of the data and reports received by \ac{iot} layer. Their role is important because they make decisions for the progress of institutions or the whole education system. To assist decision-making, top administrators can utilize smart technologies to generate more precise and accurate data. 

Children are the most important assets for \emph{parents}. Unfortunately, in developing countries, parents are not involved in the educational process. Because of their financial problems, they are also not keen in supporting their children. It is crucial to properly involve them in the teaching and learning process. Due to their passive participation, children are not properly focused in the learning process. The fog layer automatically keeps them up to date when certain conditions are met.


\begin{table*}[ht]
\centering
\caption{ Comparative analysis of IoT devices connectivity}
 \label{tab:connectummary} 
 \newcolumntype{b}{X}
\newcolumntype{s}{>{\hsize=0.4\hsize}X}
\setlength{\extrarowheight}{2pt}%
\begin{tabularx}{\textwidth}{s s s s s s} 
\toprule
\textbf{Board} &  \textbf{Network} & \textbf{Bandwidth} &  \textbf{Date Rate} & \textbf{Range} &  \textbf{Standardization} \\
\midrule
 
Bluetooth  \cite{SEI57}    &  PAN  & 1 MHz    &  1- 3 Mbit/s  & 100m   & IEEE 802.15.1   \\

Wi-Fi \cite{SEI58}  & PAN  &  20 or 40 MHz  & 10 mbps  & 100m  &  IEEE 802.11  \\

  ZigBee \cite{SEI59}  & WAN  &  2 MHz    &  250 kbps & Less than 1 km  & ZigBee alliance   \\

NB-IoT  \cite{SEI60}  &  WAN & 200 KHz    & 200 kbps  & More than 1 KM   & 3GPP   \\
\bottomrule    
\end{tabularx}
\end{table*}


\section{Research Methodology}
\label{rm}
The recent development in smart systems and a massive increase in smart devices encourage us to review the integration of smart systems and education. As per the importance of education, it is not yet integrated into smart systems (e.g,  smart cities and smart health, etc). Exploring the literature shows that only limited research articles are available on smart education. Because of these limitations, we have analyzed the literature in depth for issues in traditional education, possible smart solutions, and the challenges in smart solution implementation.

\begin{table*}[ht]
\centering
\caption{Research string used to search the related literature}
 \label{tab:ResearchString} 
 \newcolumntype{b}{X}
\newcolumntype{s}{>{\hsize=.4\hsize}X}
\newcolumntype{c}{>{\hsize=.6\hsize}X}
\setlength{\extrarowheight}{2pt}%
\begin{tabularx}{\textwidth}{s s c} 
\toprule
\textbf{Area} &  \textbf{Keywords} & \textbf{Synonyms in literature} \\
\midrule

Population  &	Education   &	Education OR Campus OR University OR Class \\
Methodology & 	Smart Devices &	Internet of Things OR IoT \\
Outcomes &	Smart Education &	Smart Education OR Smart Campus OR University \\

\bottomrule    
\end{tabularx}
\end{table*}

The emergence of the \ac{iot} and smart systems is not so earlier, however, education has been since the start of human beings. To get a broader vision of pure education, we have looked at earlier research papers, however, for the smart education implementation, we have only included the document from 2015 to 2020. We further included projects carried out to make any component of education smart (e.g, learner attendance and security systems etc).
We used Google Scholar, Scopus, IEEE Xplore, and Science Direct for searching the target papers. Google Scholar gives access to every paper published in any journal and the research libraries give access to limited high-quality papers published in attached journals and publishers.

To search the digital world, the searching string is needed and the quality of searching depends on this searching string. The search string is the keywords covering the population, methodology and outcomes. 
This research paper methodology is divided into three phases (i) The planning phase, (ii) Conducting phase, and (iii) Reporting phase. The remaining part of this section explores these phases.

\subsection{Planning the Review}
In the first phase, we planned the study protocol, searching related journals and papers, inclusion and exclusion criteria, and reporting. The planning phase covers two important aims; (i) the importance and need of the study which distinguish this study from other related studies; and (ii) developing the protocol to search the studies and inclusion and exclusion criteria. 

Recent technologies are revolutionizing the way of living and thinking. Every aspect of life is getting smarter, such as smart cities, smart health, smart agriculture, smart power plant, etc. Education is the most important domain; which includes every human being. However, the research was not done based on its importance. The assumption of 75 billion devices means 10 devices per head. Most of the time, these devices are used for entertainment, not for learning. Smart education can radically change education by using these devices to make every aspect smart.

Developing the review protocol is crucial and critical. The appropriate protocol leads us to a good review; however, the invalid protocol may lead the authors in other directions, leaving the main focus point. Therefore, emerging research questions, search strategies, and selection criteria are discussed and identified at this stage.

\subsection{Conducting the Review}
In this phase, the research is conducted as per the protocol designed in phase 1. 

 The most pivotal is the research identification and for this purpose, every research is analyzed for three important checks.

\begin{enumerate} 
\item The first one is the  \emph{population} of the research. For example, this research covers education and the \ac{iot}, therefore, the population of this article is \emph{educational} and emerging \emph{technologies}. 
\item	The second check is the \emph{methodology} or \emph{technique} which is used to get the desired outcome. In this article, smart technologies are techniques to get the desired results. 
\item	The third one is the \emph{outcome} achieved at the end of the research. In this case, the outcome is smart education. 
\end{enumerate} 

Pursuing these checks, the next most important phase is to design research questions. In this context, the research questions are;

\begin{enumerate} 
\item	How many research papers are published discussing the integration of education and smart systems?
\item	What are the main issues in the traditional education system?
\item	What are the possible uses of smart devices and techniques in smart education?
\item	What are the possible smart solutions to the issues in the traditional education system?
\item	What are the main computational and social challenges in implementing the smart education system?
\end{enumerate}

RQ1 deals with the population, RQ2 deals with methodology, and RQ3 and RQ4 deal with the outcome.


For complete strings, we combine them with  "AND", for example;

Population AND Methodology AND Outcomes. 

Now putting the related literature synonyms using OR logical operator.

\begin{quote}
\textit{(Education OR Campus OR University OR Class) AND (Internet of Things OR IoT) AND (Smart Education OR Smart Campus OR University)}
\end{quote}

 Table \ref{tab:ResearchString} shows the creation of the research string.

\begin{figure*}[h]
\begin{subfigure}{0.45\textwidth}
\includegraphics[width=1\linewidth]{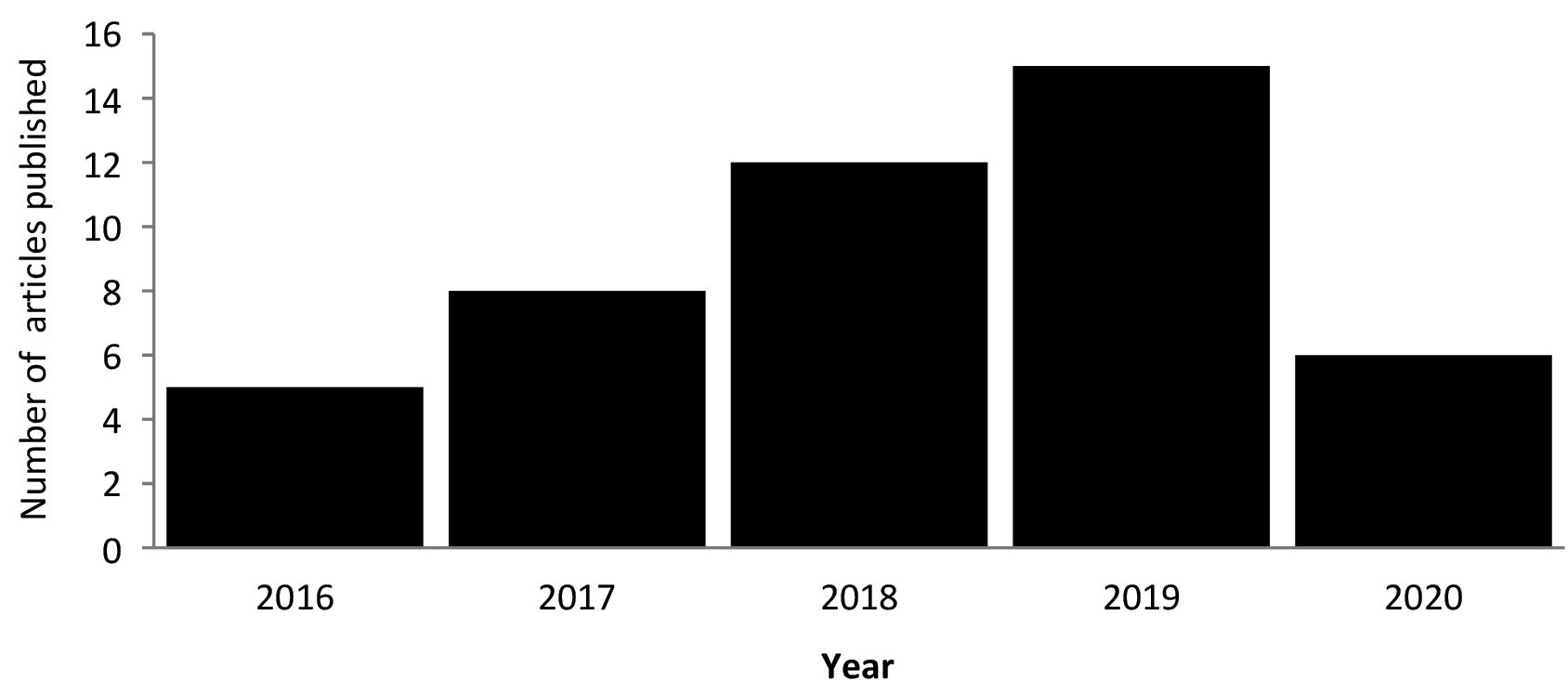} 
\caption{Number of year wise publications}
\label{fig:ywpublication}
\end{subfigure}
\begin{subfigure}{0.45\textwidth}
\includegraphics[width=1\linewidth]{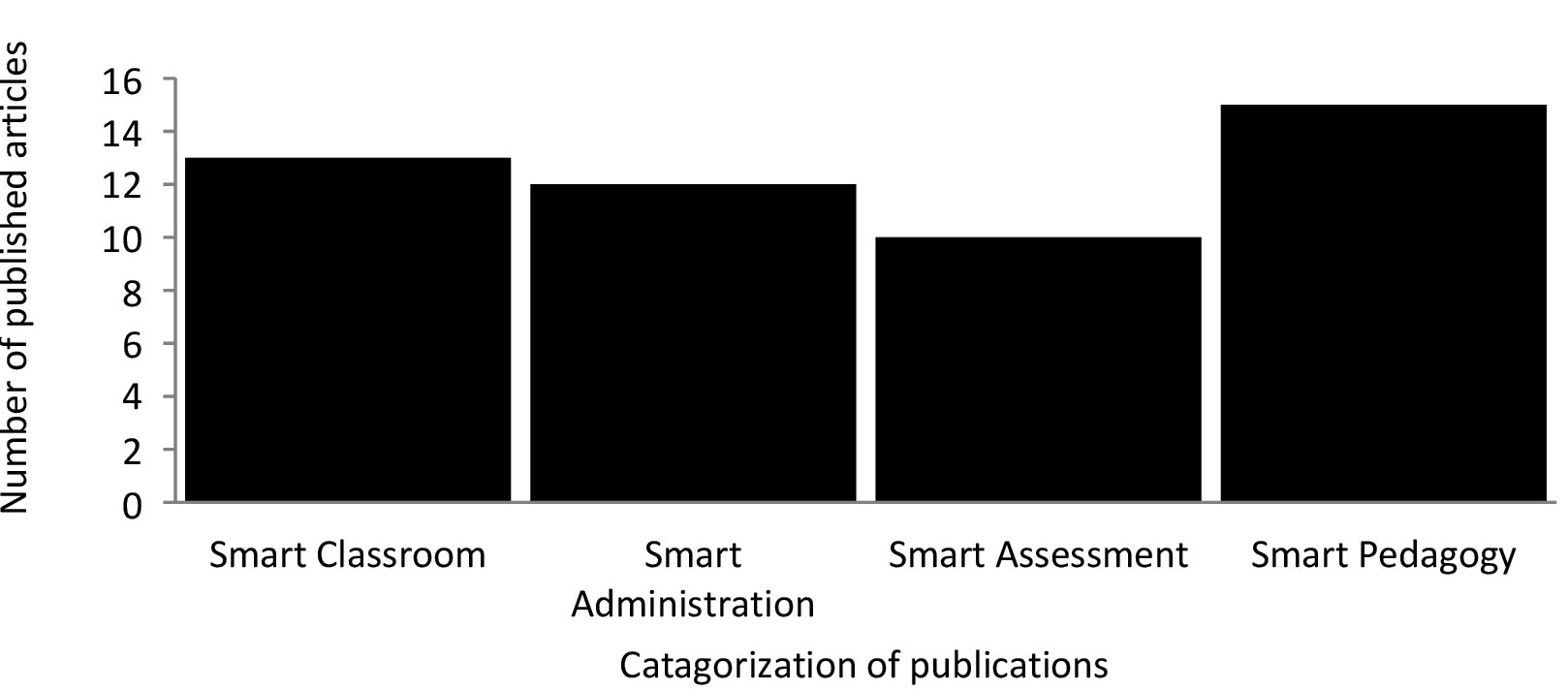}
\caption{Number of category wise publications}
\label{fig:cwpublication}
\end{subfigure}
\caption{Number of publications}
\label{fig:publication}
\end{figure*}

\subsection{Quality Assessment}
The  quality assessment of this study depends on the number of parameters. To cover this study, we have taken the following parameters to ensure the quality of the papers.

\paragraph{\textbf{Inclusion criteria:}}

\begin{enumerate} 
\item Papers covering the smart technologies (\ac{iot}, \ac{ai}, 5G, \ac{ar}, and \ac{vr} etc) integration in education in terms of smart pedagogy, smart classrooms, smart assessment and smart administration etc. 
\item A detailed description of the data in terms of devices and technology used for smart education. 
\item The presentation of the methodology and results in a proper way
\item Smart education received very limited attraction, therefore, we have included all those papers, which got one citation last year. 
\item Research articles  published since 2015.

\end{enumerate}

\paragraph{\textbf{Exclusion criteria:}}

\begin{enumerate} 
\item  The research papers discuss education or technology (single disciplinary) but do not integrate technology with education for smart education.
\item Research papers not properly presenting the results and methodology used for desired outcomes. 
\item Research papers failed to get a single citation last year. 
\item Research articles not published between 2014 and 2020.
\end{enumerate}

\begin{table*}[ht]
\centering
\caption{ Detail summary of year wise  publications since 2015}
 \label{tab:ywpublication} 
\newcolumntype{b}{X}
\newcolumntype{s}{>{\hsize=0.4\hsize}X}
\newcolumntype{a}{>{\hsize=1.5\hsize}X}
\newcolumntype{c}{>{\hsize=3.2\hsize}X}
\setlength{\extrarowheight}{2pt}%
\begin{tabularx}{\textwidth}{a c s s s s s s s s} 
\toprule

\textbf{Category} &  \textbf{Subcategory} &  \textbf{2015} & \textbf{2016} &  \textbf{2017} &  \textbf{2018}  &  \textbf{2019}  &  \textbf{2020}  &  \textbf{Total}   \\
\midrule
 
Smart Classroom & Virtual Classroom, MOOC, Smart LMS, Smart Collaboration & 1&1 &2 &3 &4&2&13 \\

Smart Administration & Smart Attendance, Smart Planning, Smart Portfolios, Smart Reports, Smart Security& 2&	1&	3&	2&	3&	1&	12 \\

Smart Assessment & Smart Question Bank, Smart Marking, Smart Observation, Smart Applications &0&	2&	2&	2&	3&	1&	10 \\

Smart Pedagogy & Smart Engagement, Flipped Classroom, Smart Activities, Personalized Learning, Smart Lesson Plan &1&	1&	1&	5&	5&	2&	15 \\

\textbf{Total}  & & 4&	5&	8&	12&	15&	6&	50 \\

\bottomrule    
\end{tabularx}
\end{table*}

\subsection{Reporting the review}
In the last stage, meaningful papers covering the keywords and research questions were extracted and presented in this study. The success of the review totally depends on the way how the final review is presented in the paper. Table   \ref{tab:ywpublication}   and Fig.  \ref{fig:ywpublication}  show the year wise  and Table    \ref{tab:cwpublication} and Fig. \ref{fig:cwpublication} show the category-wise publications since 2015.

 \begin{figure*}[h]
 \center
	\includegraphics[scale =.4]{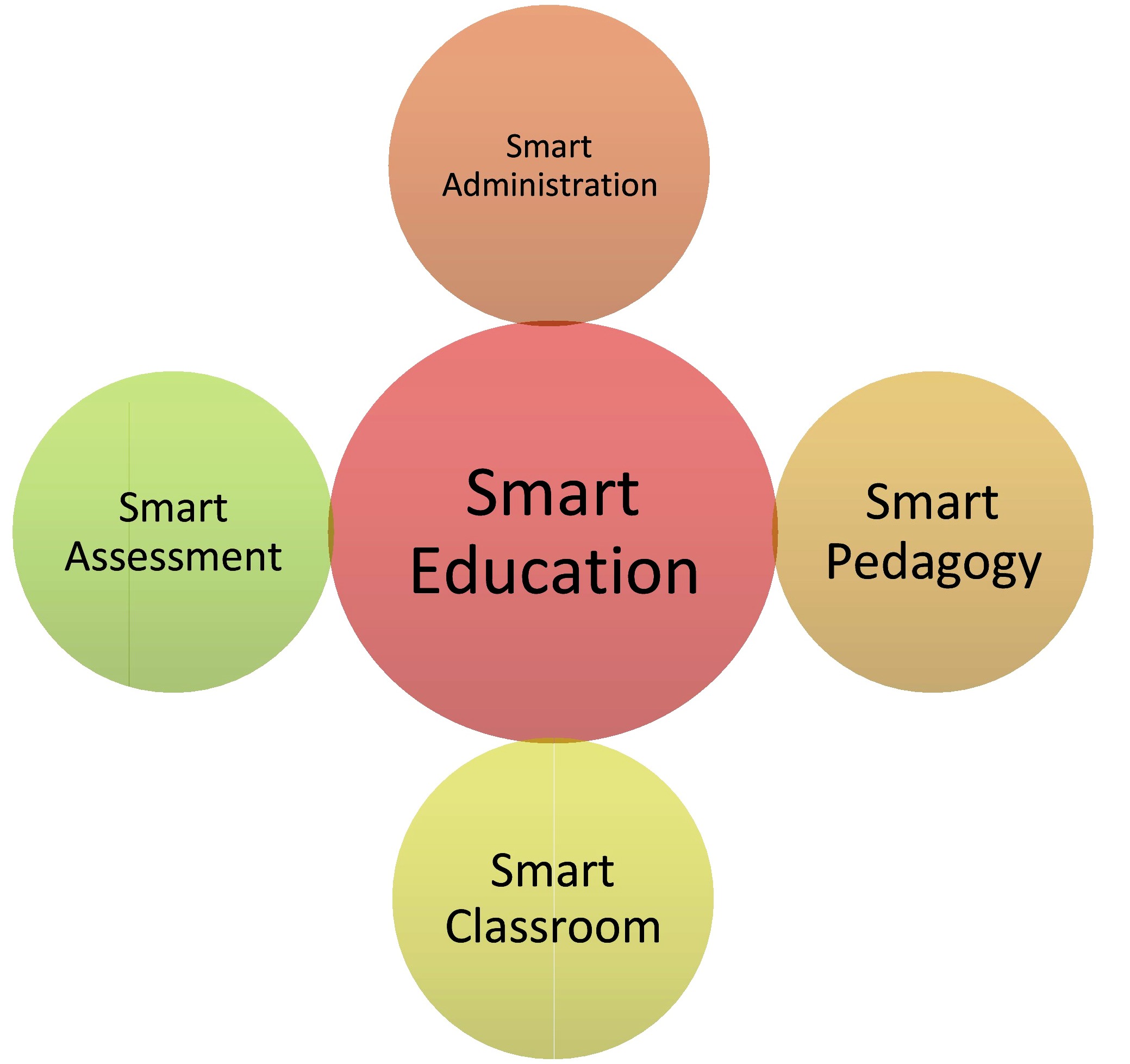}
	\caption{Classification of smart education}
	\label{fig:seclassification}
\end{figure*}

\section{Literature classification}
\label{literature}
\ac{iot}, \ac{ai}, and 5G are changing the world and way of living. It has its share in all fields of life. Educational institutions are part of everyone's life and  as per their importance, it has not  been investigated to make them smart. \ac{iot} in education requires further research. This section has reviewed and classified (as shown in Fig  \ref{fig:seclassification}) the related investigations and summarized the findings in  Tables \ref{tab:ssclassroom}, \ref{tab:ssadministration}, \ref{tab:sassessment}, and \ref{tab:sspedagogy}.

\subsection{Smart classroom}
With the development of technology, teaching procedures changed. Now technology is used to engage the learners and make the classroom productive.  Smart classroom uses  smart technology to enhance the learning environment. Table \ref{tab:ssclassroom} shows a detailed review of the related studies in the smart classroom.

In most regions, cultures, and religions;  co-education is not liked (male and female learners are not liked to sit in one classroom). In this social context, \citet{SEI50} investigated this issue and proposed a framework where male teachers teach male  and female learners virtually attend  lectures on a laptop, tablet, or mobile. 
\citet{SEI03} further explored the virtual learning environment and the role of \ac{iot} in it.  They used the fog and cloud computing layers to process large amounts of data produced by different devices in virtual learning.  Authors in \cite{SEI102} proposed a two-way digital teaching method. The learner computers and tablets show the learner agents, similarly, the teacher's tablet shows the teacher agent.  The main challenges with virtual classrooms are that teachers are not able to effectively monitor the learners. Furthermore, students learn with different teaching methodologies, however, the virtual classroom does not support allowing different methodologies \cite{SEI108}. 

Classroom and seating management is also a challenge in educational institutions.  Authors in  \cite{SEI100} proposed a framework to schedule the classroom for students.
This study was further pursued by \citet{SEI101} and proposed a smart framework for educational institutions to smartly manage the attendance and seating layout. 
Learning increases with flexible classrooms and decreases if the classroom environment is not appropriate, especially in temperature.
\citet{SEI18}, investigated this further by proposing a flexible classroom. In this classroom, learners can easily change their location and directions for better interaction. The instructor taught  Introduction to Electronics Circuits (IEC) as a traditional class for one year and then a flexible classroom for the next year. They found that the flexible classroom results were better than the traditional classroom. Pursuing the same concept,  \citet{SEI55} focused on the underutilized classrooms on a university campus due to the gap between enrollment and attendance. They used \ac{iot} sensors to efficiently use classrooms in real-time. With  \ac{ai} (to predict the presence and optimum performance assignment of classrooms), they  minimized the waste of space.
\ac{ai} faces different challenges in practical life. In the above case,  it needs a powerful algorithm to detect the seating space in the room. Furthermore, there are also chances to make mistakes in reading and allotment \cite{SEI109}.

In learning-centred classrooms, learners are engaged in creative thinking and learning.
\citet{SEI19} discussed the effectiveness of a learner-centred teaching approach. He explored the barriers towards learner-centred such as learners' significance of e-learning materials, the extra time required for e-learning material, learners' resistance to taking an active role, and teacher resistance towards learner-centred approaches.  \citet{SEI22},  further explored the use of smartphones to increase learner engagement. Similarly, authors in \citet{SEI103}  developed a real-time feedback system named as MTFeedback. This automatically monitors the collaboration of smart groups and notifies the teacher in real-time.

Open education is seen as an evolving teaching method. \ac{mooc} is the new trend in open education and the world-famous universities offer \ac{mooc} courses. The integration of \ac{ar} and \ac{vr} with \ac{mooc} is also interesting, attracting the masses there. The authors of \citet{SEI93} evaluate how \ac{mooc} lessons work. The result shows that it not only improves students' knowledge but also makes lessons easy and enjoyable. 

One of the dominant concerns with large classes is that students become shy and they cannot dare to ask questions or take part in the classroom's communication. It is a universally accepted truth that such students pay a high price for shyness in future.  To overcome this issue and increase the collaboration among students, the authors in  \cite{SEI94} developed a framework where learners collaborate with each other to complete and discuss a task. This increases collaboration among students.

\subsection{Smart assessment}
Assessment is a crucial part of teaching and learning; used to assess learners, teachers, and syllabus performance. The smart assessment uses ICT technology to automate and transparent this process. Table \ref{tab:sassessment} shows a detailed review of the related studies on smart assessment. 

For transparent reports, \citet{SEI09} investigated the mobile-based assessment. They developed an application having a teacher and learner interface. Teachers develop tests for learners and learners solve these tests on mobile or tablets. The papers are automatically marked, saving time and cost.   \citet{SEI104} developed an application which facilitates teachers to create assessment activities for students to create high-order thinking activities. 
Similarly,  \citet{SEI89} developed a chatbot programming application for learners' assessment. This application aimed to increase students' engagement and task completion. The case study results show that Chatbot users were more engaged than the other learners.  In another study,  \citet{SEI95} proposed an application where teachers create different types of activities for learners and learners solve these activities on their mobile phones and  tablets. The application automatically calculates the status of the learner in any subject. 

In another case study, \citet{SEI88} combined the flapped classroom and continues assessment. The results show a great inclination. The dropout rate was reduced and the learners passed both the practical and theoretical portions. The most important of this model is that the learner accepts this combination for learning and assessing. \citet{SEI92} developed an application which helps learners to improve their pronunciation and assess their learning at the end of the class.

\subsection{Smart administration}
Smart administration is the use of a sensor-enabled environment, where it reads  the  physical changes and automatic decisions are made. Smart administration has been under discussion since  2010. Table \ref{tab:ssadministration} shows a detailed review of the related studies on smart administration.

Attendance irregularities badly affect educational institutions' performance. However, now biometric, \ac{rfid} and face detection technologies are used to automatically manage the attendance \cite{SEI99} . \citet{SEI04} used the \ac{rfid} for learner attendance system. \ac{rfid}  was installed on Micro-controller. Each learner receives integrated cards with \ac{rfid} technology. When learners enter the classroom, their attendance is automatically marked \cite{SEI174} .

\onecolumn
\begin{landscape}
\begin{longtable}{p{2cm}p{0.5cm}p{2cm}p{6cm}p{3cm}p{5cm}}
\caption{Detail summary of studies related to smart classroom}
\label{tab:ssclassroom} \\

\toprule
\textbf{Paper} & \textbf{Year} &   \textbf{Idea} &   \textbf{Findings} & \textbf{Device / Technology Used} &   \textbf{Limitations} \\ 
\midrule
\endfirsthead
\multicolumn{2}{l}{\footnotesize\itshape\tablename~\thetable:
continued from previous page} \\
\toprule
\textbf{Paper} & \textbf{Year} &   \textbf{Idea} &   \textbf{Findings} & \textbf{Device / Technology Used} &   \textbf{Limitations} \\

\midrule
\endhead
\midrule
\multicolumn{2}{r}{\footnotesize\itshape\tablename~\thetable:
Continued on next page} \\
\endfoot
\bottomrule
\multicolumn{2}{r}{\footnotesize\itshape\tablename~\thetable:
It ends from the previous page.} \\
\endlastfoot

\citet{SEI50} & \citeyear{SEI50} & Virtual classroom &   In some regions (social and religious context) boys and girls are not like to sit together. The authors  proposed a framework where male teachers teach the male learners and female learners virtually attend lectures on laptop, tablet or mobile.  & Fog computing, cloud computing, PDAs, and laptops &   In virtual classes, the main issue is to engage and monitor the learners. The authors did not discuss, how the learners are observed and engaged in virtual classrooms  \cite{SEI108}. \\ 

\citet{SEI55} & \citeyear{SEI55} & Smart classroom &   Authors used \ac{iot} sensors to efficiently use the classrooms in real-time. With  AI (to predict the presence and optimum assignment of classrooms), they minimize the waste of space.  & \ac{ai}, Fog computing,  PDAs, and Laptop &  This automatically shows the used and available spaces, however, the case of lack of building still exists. Furthermore,  \ac{iot} may make mistakes in reading the space \cite{SEI109}. \\ 

\citet{SEI03} & \citeyear{SEI03} & Virtual classroom &  Authors proposed a big data and fog-based e-learning environment, where learners virtually attend the classes. With virtual classrooms, a personalized learning environment was created for students. & Fog computing, cloud computing, PDAs, \ac{ai} &  In virtual classes, the main problem is to engage and monitor the learners. Students learn with different methodologies. The major lacks in virtual classes are that this framework does not support multi-teaching methodologies \cite{SEI108}.  \\ 

\citet{SEI05} & \citeyear{SEI05}  & Smart classroom & This is a case study replacing the lecture method with the smart classroom. They proposed  active learning classrooms to actively engage learners in classes. The results of this case study were promising.  & Smartboards, PDAs  & This research study is limited to engineering students studying in the Middle East     \\

\citet{SEI93} &  \citeyear{SEI93} & Smart classroom & This is a study case to evaluate the working and performance of \ac{mooc} classes.   &  Mobiles phones, laptop etc &  Such like applications helps the adults. This study misses discussing the solution to use it by kids.  \\

\citet{SEI100} &  \citeyear{SEI100} & Smart classroom  & An algorithm was developed to sense the capacity of the classroom and allow this  to the students as per the morning attendance &  Cameras, \ac{ai}, Mobile phones etc & This uses the \ac{ai} which is not so mature to correctly sense the classroom environment \cite{SEI109}.  \\

\citet{SEI101} &  \citeyear{SEI101} & Smart Classroom &Authors proposed a smart framework for educational institutions to smartly manage the attendance and seating layout &  Cameras, \ac{ai}, Mobile phones etc & The major challenge with this framework is the use of AI to detect the seating capacity \cite{SEI109}. \\

\citet{SEI102} &  \citeyear{SEI102} & Behaviour recognition  & Authors proposed a teaching method using the teacher and students agents. This technique is used to monitor the learner behaviour on teacher tablets &  Cameras, \ac{ai}, Mobile phones etc &  This uses the \ac{ai}  for behavioural recognition, however, the   \ac{ai} is not so mature to perfectly sense the classroom environment \cite{SEI109}.      \\

\bottomrule
\end{longtable}

\end{landscape}
\twocolumn

\onecolumn
\begin{landscape}
\begin{longtable}{p{2cm}p{0.5cm}p{2cm}p{6cm}p{3cm}p{5cm}}
\caption{Detail summary of studies related to smart assessment}
\label{tab:sassessment} \\
\toprule
\textbf{Paper} & \textbf{Year} &   \textbf{Idea} &   \textbf{Findings} & \textbf{Device / Technology Used} &   \textbf{Limitations} \\ 
\midrule
\endfirsthead
\multicolumn{2}{l}{\footnotesize\itshape\tablename~\thetable:
continued from previous page} \\
\toprule
\textbf{Paper} & \textbf{Year} &   \textbf{Idea} &   \textbf{Findings} & \textbf{Device / Technology Used} &   \textbf{Limitations} \\
\midrule
\endhead
\midrule
\multicolumn{2}{r}{\footnotesize\itshape\tablename~\thetable:
Continued on next page} \\
\endfoot
\bottomrule
\multicolumn{2}{r}{\footnotesize\itshape\tablename~\thetable:
It ends from the previous page.} \\
\endlastfoot

\citet{SEI09} & \citeyear{SEI09}  & Smart Assessment &  Authors developed an application for mobile-based learners assessment. They utilized a NodeJS server that communicated with CouchDB, an MQTT broker, and an  analytic engine to run this application.    &  PDAs, mobiles phones, database, cloud computing  & This study did not discuss the  solution for  kids appearing in the mobile-based assessment \cite{SEI111}.  \\

\citet{SEI95} &  \citeyear{SEI95} & Smart Assessment & Proposed an application where teachers create different types of activities for learners and learners solve these activities on their mobile phones tablets & Social media, mobiles phones, laptops etc  & Creating activities is time taking task. Furthermore, most of the teachers would not be able to create activities.  \\

\citet{SEI92} &  \citeyear{SEI92} & Smart Assessment & Proposed an application  for learner pronunciation improvement and analysis &Mobiles phones, laptop etc  &  Such like applications helps the adults. This study misses discussing the solution to use it by kids.  \\

\citet{SEI89} &  \citeyear{SEI89} & Smart Assessment  & Proposed a  Chatbot application. Student program it to learn the basic concept of computer programming. It provides a formative assessment to the students.  &  laptop, mobile phones &   This application is missing such like application used in kids. \\
  
\citet{SEI88} &  \citeyear{SEI88} & Smart Assessment  & They developed an online tool which supports teachers in paper-based exams.  The tool's result shows  transparency, quick response and error-free results. &  laptop, mobile phones & Such applications help adults. This study misses discussing the solution to use it by kids.  \\

\bottomrule
\end{longtable}

\end{landscape}

\twocolumn

The smart-university  concept was discussed by \citet{SEI97}. They proposed an \ac{iot} setup to automatically manage the university, such as smart lighting, smart parking, smart tracking, smart inventory etc. 
Similarly, authors in \cite{SEI07}, explored the potential  benefits of \ac{iot} in education. This study covered cost reduction, performance improvement, new revenue generation,  customer experience enhancement, differentiated service creation, and  the power of \ac{iot} in the classroom.  
 
 The teaching process is progressing towards digitization. 
\citet{SEI01} investigated digitization and  discussed the key impacts and challenges in this process. It was expected that learners, teachers, curriculum, and institutions would  possibly take benefit from this conversion.  They further  discussed the role of leadership, culture, technology, and the methodology used to do the conversion.   \citet{SEI23} further investigated the use of ICT in the education system. 
\citet{SEI25} discussed smart education to solve traditional educational issues. They focused on personalized and seamless learning. 
\citet{SEI26} worked on NEdNet (National Education Network) in Thailand. Their finding shows that ICT increased the quality of education. 

Internet of Everything (IoE) extends the concept of \ac{iot} to machine-to-machine communication. 
\citet{SEI02} discussed the  use of  the Internet of Everything (IoE) in education. According to their statistics, about 50 billion devices will be connected to the internet by 2025. The authors discussed  people, processes, data, and things as key pillars of IoE. Key factors to successfully implement the IoE in education are security, data integrity, and educational  policies.

Security is a critical issue in institutions. Records show that hundreds of learners were killed in institutions due to the lack of security measures. 
\citet{SEI27} proposed the Smart Security Framework (SSF) for educational institutions. This work was motivated by the recent attacks on the Pakistan and USA educational institutions. Smart devices are used to read the physical environment. The data is forwarded  to fog computing. The fog layer generates a notification in case of any threat. Higher-level authorities are informed in case of serious threats.

\ac{ai} is playing a big role in society. \citet{SEI67} proposed an idea to teach medical students using AI. They further proposed to use \ac{ai} devices to directly interact with patients. 
\citet{SEI33} proposed a framework for smart decisions in the school. They discussed the educational institutions' issues which can be handled through \ac{iot} as in other fields such as health,  agriculture, etc.   \citet{SEI68} proposed an idea which uses AI techniques and cognitive science to try and understand the nature of learning, teaching, and creating systems to help teachers to apply new skills or understand new concepts.

\subsection{Smart pedagogy}
Pedagogy is the heart of the teaching process and smart pedagogy relates to the methodologies used in teaching and learning. As a science, it searches for new ways to enhance student learning. The most important development in pedagogy was seen in 1956 when the learning taxonomy was presented by Bloom \citet{SEI113}.  Table \ref{tab:sspedagogy} shows a detailed review of the related studies on smart assessment.

The first detail work was carried out by \citet{SEI105} developing a complete structure for smart learning and assessment. 
\citet{SEI20}, explored the Intelligence of Learning Things (IoLT). The proposed framework is based on collaborative learning, middleware, and personalized learning. The available technologies were used to get cost-effective solutions for developing countries. The result analysis shows that the \ac{iot} has a big impact on cost, performance, and safety.  Similarly, the authors in  \citet{SEI98} proposed an indoor learning framework for deep learning through the working of the algorithm.

Smart glasses have key potential in the education system. 
\citet{SEI08} explored its potential use  in teaching and learning. They found that smart glasses have the key potential for teleconferencing, telemetering, online evaluation of teachers and trainers, listener experience, real-time teaching, etc. 

The use of mobile phones is rapidly increasing. This may be used in smart learning. 
\citet{SEI21} reported a research project in a university, where every employee voluntarily brings their own devices. They used this project to measure the quality enhancement of assessment using smart devices. They worked on  Bring Your Own Device (BYOD) and six Cs (i.e, connect, communicate, collaborate, curate, create, and coordinate) to analyze the results. The results proposed that instead of the above BYOD and 6 Cs, there are a lot of other needs to convert the institutions to smart institutions. Similarly, \citet{SEI92} developed an application to enhance language pronunciation improvement and learner engagement.

Passive classes waste the learners' time. The flipped classroom is a new teaching methodology where learners watch videos at home and do activities at school under teacher supervision. 
\citet{SEI05} tried to replace the traditional passive classes with active learning. 
 In the same way, \citet{SEI06} further explored the flipped classroom to deliver classes online with smart devices and these lesson activities are carried out in class. Instead of face-to-face teaching, learners attend classes at home using Learning Management System (LMS). During the face-to-face session, related activities are carried out to keep them engaged in productive activities. Learners access all material at  home. As a case study, \citet{SEI48} used a flipped classroom for the subject of the web server for master classes and worked on the improvement of assessment. The results show that learners' response was optimum comparatively to previous years' results.

 \onecolumn
 \begin{landscape}

\begin{longtable}{p{2cm}p{0.5cm}p{2cm}p{6cm}p{3cm}p{5cm}}
\caption{Detail summary of  smart administration}
\label{tab:ssadministration} \\

\toprule
\textbf{Paper} & \textbf{Year} &   \textbf{Idea} &   \textbf{Findings} & \textbf{Device / Technology Used} &   \textbf{Limitations} \\ 
\midrule

\endfirsthead
\multicolumn{2}{l}{\footnotesize\itshape\tablename~\thetable:
continued from previous page} \\
\toprule
\textbf{Paper} & \textbf{Year} &   \textbf{Idea} &   \textbf{Findings} & \textbf{Device / Technology Used} &   \textbf{Limitations} \\

\midrule
\endhead

\midrule
\multicolumn{2}{r}{\footnotesize\itshape\tablename~\thetable:
Continued on next page} \\
\endfoot
\bottomrule
\multicolumn{2}{r}{\footnotesize\itshape\tablename~\thetable:
It ends on the previous page.} \\
\endlastfoot

\citet{SEI04} & \citeyear{SEI04} & Smart attendance &  Attendance irregularity is the main concern of educational institutions. In this article,  a smart attendance and notification  system was proposed for learners to efficiently manage attendance. &  \ac{mcu}, \ac{rfid}, PDAs, Mobile phones  & The main concern with smart attendance is that it does not work in a remote areas, where internet connection problems exist \cite{SEI43}.    \\

\citet{SEI27} & \citeyear{SEI27}  & Smart security &  Authors proposed a smart security framework for educational institutions for learners' and teachers' safety. They used cloud computing, fog computing, \ac{iot} and various sensors. The results of this study were promising  & \ac{mcu}, cameras, sensors, fog and cloud computing. & This project was simulated and not tested in a real physical environment. Therefore, it may not be trusted in real environments. \\

\citet{SEI33} & \citeyear{SEI33}  &  Smart administration &  Authors proposed a framework for smart decisions in educational institutions. They used the \ac{iot} and related technologies to efficiently make decisions for teaching and learning.   & Cameras, mobile, tablets, fog and cloud & Smart administrations are usually resisted by employees. The authors did not discuss how to overcome these challenges \cite{SEI82}.     \\

\citet{SEI68} & \citeyear{SEI68}  & Smart learning &  They developed a framework using AI and cognitive science to understand the nature of learning and teaching and according to this analysis, teachers apply new skills for teaching. & Cameras, mobile, tablets, fog and cloud & The major challenge with this is the use of AI because the AI is not so mature to trust its results \cite{SEI109}.    \\

\citet{SEI94} &  \citeyear{SEI94} & Smart collaboration &  Developed a framework where learners collaborate with each other to complete and discuss a task. This increases collaboration among students. & Social media, mobiles phones, laptops etc & online collaboration have issues such as network failure etc \cite{SEI110}. \\

\citet{SEI97} &  \citeyear{SEI97} & Smart university &  Authors proposed an \ac{iot} setup to automatically manage the university such as smart lighting, smart parking, smart tracking and smart inventory etc. & AI, sensors and cameras & The use of AI faces many challenges for correctly configuring the things \cite{SEI109}.  \\

\bottomrule
\end{longtable}

\end{landscape}

\begin{landscape}

\begin{longtable}{p{2cm}p{0.5cm}p{2cm}p{6cm}p{3cm}p{5cm}}
\caption{Detail summary of smart pedagogy}
\label{tab:sspedagogy} \\

\toprule
\textbf{Paper} & \textbf{Year} &   \textbf{Idea} &   \textbf{Findings} & \textbf{Device / Technology Used} &   \textbf{Limitations} \\ 
\midrule

\endfirsthead
\multicolumn{2}{l}{\footnotesize\itshape\tablename~\thetable:
continued from the previous page} \\
\toprule
\textbf{Paper} & \textbf{Year} &   \textbf{Idea} &   \textbf{Findings} & \textbf{Device / Technology Used} &   \textbf{Limitations} \\

\midrule
\endhead

\midrule
\multicolumn{2}{r}{\footnotesize\itshape\tablename~\thetable:
Continued on next page} \\
\endfoot
\bottomrule
\multicolumn{2}{r}{\footnotesize\itshape\tablename~\thetable:
It ends on the previous page.} \\
\endlastfoot

\citet{SEI08} & \citeyear{SEI08} & Smart pedagogy &  Authors investigated the use  of smart glasses in the teaching and learning process. This case study shows that smart glasses, virtual and augmented reality have enough potential to enhance the teaching and learning process & Smart glasses, Virtual Reality, Augmented Reality &  Smart glasses help to improve learning by presenting the topics in  3D concepts, however, these are expensive and every department cannot support to buy and use \cite{SEI112} .  \\

\citet{SEI67}  & \citeyear{SEI67}   & Smart pedagogy &  They proposed an idea to teach medical students using AI. They further proposed to use AI devices to directly interact with patients & Artificial Intelligence, cameras, mobile, tablets, fog, and cloud   & The project is limited to medical students. Furthermore, the use of \ac{ai} in education is complicated. The authors did not explore, how they employed the \ac{ai} in teaching and learning \cite{SEI114}.    \\

\citet{SEI68} & \citeyear{SEI67}  & Smart pedagogy &  They proposed an idea which uses \ac{ai} techniques and cognitive science to try and understand the nature of learning and teaching & \ac{ai}, cameras, mobile, tablets, fog, and cloud   & The use of \ac{ai} in education is complicated. The authors did not explore, how they employed the \ac{ai} in teaching and learning \cite{SEI109}.    \\

\citet{SEI48} & \citeyear{SEI48}  & Flipped classroom &   Authors used flipped classroom (Learners watch recorded videos at home and perform activities at school under teacher supervision) for one of the master classes. The result shows that learners' response was good compared to previous years' results.   & Cameras, mobile, tablets, fog, and cloud & This system does not guarantee that learners watch the recorded video at home and come to school perfectly prepared for the activities \cite{SEI115}.   \\

\citet{SEI06} & \citeyear{SEI06} & Flipped classroom&  Flipped classrooms is the emerging teaching methodology. They used Classroom Response Systems (CRS) in which learners respond to the classroom's activities on mobiles or PDAs.   & Mobiles phones, PDAs, cameras, cloud computing  &  As discussed, the challenge with the flipped classroom is that usually students do not watch videos at home and come to school unprepared.   \\

\citet{SEI10} & \citeyear{SEI10}  & Flipped classroom &  Flipped classrooms is the emerging teaching methodology. The authors in this paper worked to combine the flipped classroom and traditional classrooms. The result of the study was promising in terms of critical thinking and knowing.  & Cameras, mobile phones, tablets, fog, and cloud  & Flipped classrooms and traditional classrooms are two different ways of teaching. This article lacks a suitable framework to combine the two in one class. \\

\bottomrule
\end{longtable}

\end{landscape}

\twocolumn

\citet{SEI10}  further explored the flipped classrooms for engineering learners which enhanced the creativity and engagement of the learners. 
\citet{SEI17}, carried out experimentation on electrical engineering classes by using the traditional class and flipped classrooms. Separated two classes of 20, 20 learners were created. In the flipped classroom, video lecture was the primary source of teaching. The comparative result shows that overall the flipped class learners' performance was better than the traditional one.  

\section{The key enabling technologies for smart institutions}
\label{uses}


This section discusses the key enabling technologies for smart institutions.

\subsection{Smart devices}
Different types of smart devices and sensors are used to update data on servers working on the fog layer. These are devices installed in the institution, working automatically or palmtop operated by teachers, learners, administrators, or parents (e.g, sensors, cameras, smart board, tablet and laptop etc). Sensors read the physical or chemical changes in something. 

\paragraph{\textbf{\ac{rfid}:}}
It automatically identifies the objects by the frequency attached to them. In institutions,  \ac{rfid} technology is used in smart cards (learners and staff) to make their attendance in classes.  \ac{rfid} readers are installed at the entrance to the class. When a teacher or learner enters the class, attendance is automatically marked on the database with the entrance time. \ac{rfid} saves time for the classes, which is usually wasted in taking roll calls. Furthermore, due to exact timing reporting, teachers and learners  attend the class on their proper time which is usually not seen in under-developed countries \cite{SEI64}. 

\paragraph{\textbf{Smartboard:}}
Learners forget what they listen, they learn what they watch.  The interactive smart board helps them in their learning. They watch the practical doings instead of only listening to theoretical lectures by teachers.  Smartboard is the iconic hub of modern classes. This uses modern teaching methodology to keep learners engaged. Teachers may display their own models or search for any other one on the cloud or internet. Interactive boards are easy to use and anyone may use them.  Game-based activities may easily be used to teach children to make their interest in learning \cite{SEI148, SEF116}.

\paragraph{\textbf{\ac{vr} and \ac{ar}:}}  \ac{vr} and \ac{ar} are becoming a part of teaching and learning. This helps the teachers to teach the concept of a three-dimensional environment just like real physical objects. This simulates the real environment for learners \cite{SEI91}.  Studies show that it develops creativity, promotes visual learning, and increases students' attendance \cite{SEI205}.

\begin{table*}[ht]
\centering
\caption{ Category wise publications since 2015}
 \label{tab:cwpublication} 
 \newcolumntype{b}{X}
\newcolumntype{s}{>{\hsize=0.26\hsize}X}
\newcolumntype{c}{>{\hsize=0.5\hsize}X}
\newcolumntype{a}{>{\hsize=0.2\hsize}X}
\setlength{\extrarowheight}{5pt}%
\begin{tabularx}{\textwidth}{a s  c} 
\toprule

\textbf{Category} &  \textbf{Subcategories} &   \textbf{Key studies}  \\
\midrule

Smart classroom & Virtual Classroom, MOOC, Smart LMS, Smart Collaboration    & \citet{SEI50}, \citet{SEI03}, \citet{SEI102}, \citet{SEI100},  \citet{SEI101} , \citet{SEI18},  \citet{SEI55},  \citet{SEI19},  \citet{SEI22},  \citet{SEI103} ,  \citet{SEI93}, \citet{SEI94}, \citet{SEI149},   \\

Smart Administration &	Smart Attendance, Smart Planning, Smart Portfolios, Smart Reports, Smart Security   &  \citet{SEI97}, \citet{SEI99},  \citet{SEI04},  \citet{SEI07}, \citet{SEI01} ,   \citet{SEI23},  \citet{SEI25} , \citet{SEI26},  \citet{SEI02},  \citet{SEI27},  \citet{SEI67},  \citet{SEI33},  \citet{SEI68},  \citet{SEI09},  \citet{SEI104},  \citet{SEI89},  \citet{SEI95},   \citet{SEI88},   \\

Smart Assessment &	Smart Question Bank, Smart Marking, Smart Observation, Smart Applications   &  \citet{SEI09}, \citet{SEI104}, \citet{SEI89},     \citet{SEI95},  \citet{SEI88} \\

Smart Pedagogy &	Smart Engagement, Flipped Classroom, Smart Activities, Personalized Learning, Smart Lesson Plan   &  \citet{SEI105},   \citet{SEI98},  \citet{SEI08},  \citet{SEI21},  \citet{SEI92},  \citet{SEI05}, \citet{SEI06} , \citet{SEI48},  \citet{SEI10}, \citet{SEI17},  \\

\bottomrule    
\end{tabularx}
\end{table*}


\begin{table*}[ht]
\centering
\caption{ Possible uses of smart  devices in educational institutions (Part-I)}
 \label{tab:npf1} 
 \newcolumntype{b}{X}
\newcolumntype{s}{>{\hsize=0.2\hsize}X}
\setlength{\extrarowheight}{5pt}%
\begin{tabularx}{\textwidth}{s b} 
\toprule
\textbf{Smart Devices} &  \textbf{Possible Use in Smart Education}\\
\midrule

\ac{mcu} / Smart Circuit Board  & \ac{mcu} is a single board computer, working as a hub for \ac{iot} devices. This is used to connect all devices and make decisions as per the sensors' input. A number of \ac{mcu} models are available, which are selected as per requirements.   In educational institutions, this may be used to collect data from different sensors and smart devices \\

\ac{rfid} & \ac{rfid} chip is embedded in the card, which is identified through the radio frequency.  This can help in the automatic attendance of learners and teachers. Instead of wasting time on manual attendance, \ac{rfid}  marks the attendance automatically. This will optimize the teachers' and learners' responses, performance, and behaviour \\
 
 Smart \& interactive boards  & Learners forget what they listen, they learn what they watch.  The interactive smart board helps them in their learning. They may watch the practical doings instead of only listening to theoretical lectures by the teachers. This may increase the interest and creative engagement of learners \\
 
Smartphones and PDAs  &  The use of smartphones are massively increasing and nearly every person has a mobile.  Smartphones may easily be used for teaching and learning by teachers, learners, and parents. Flip classrooms, which are becoming popular teaching methodologies depend on smartphones. \\
  
Sensors \& actuators  &  Sensors and actuators read the physical change in anything and actuators work and control according to the directions of the central system. These sensors may be used to  monitor and sense the intuitions for different parameters in real-time, for example, the temperature, motion detection, security, etc.  \\
  
 Cameras \&  security cameras & Cameras are used for monitoring and security purpose. The camera's presence changes human behaviours. This may be used in face detection attendance. In the case of flipped classrooms, they may be used by teachers to record lectures for learners.  \\
 
 \bottomrule    
\end{tabularx}
\end{table*}

\begin{table*}[ht]
\centering
\caption{ Possible use of smart  devices in educational institutions (Part-II)}
 \label{tab:npf2} 
 \newcolumntype{b}{X}
\newcolumntype{s}{>{\hsize=0.2\hsize}X}
\setlength{\extrarowheight}{5pt}%
\begin{tabularx}{\textwidth}{@{}s b@{}} 
\toprule
\textbf{Smart Devices} &  \textbf{Possible Uses}\\
\midrule

\ac{vr} and \ac{ar} & \ac{vr} and \ac{ar} is becoming a part of teaching and learning. This helps the teachers to teach the concept in a three-dimensional environment just like real physical objects. This simulates the real environment for learners. \\
  
 Smart speakers \& smart microphones & Smart speakers have different connectivity options. They may be connected by Wi-Fi, Bluetooth, or direct IP access.  This may be used for announcement and timetable management. The smart microphone may be used in classes with \ac{vr}  capabilities to automatically update the home assignment \cite{SEI65}.   \\

Alexia  & Alexia is a smart voice assistant. It uses \ac{ai} and Voice Recognition (VR) to interpret voice commands. This put great effort into educational institution automation. Instead of using notification systems, Alexia may verbally explain the situation. \\

Smart glasses   & Majorly, smart glasses are used in games and medical activities. This may be used in educational institutions to enhance the teaching and learning process. The smart glass has a good potential for telemetry and teleconferencing \cite{SEI51}.  \\

 Biometrics \&  face detection  & Manual attendance is an issue and time-consuming. It is also not secure. Biometric and face detection technologies may be used in the attendance system. This can help to keep the exact arrival and departure information. \\

Scan-marker  &  Scan marker is a device which scans the text and converts it into different languages. This may be used in reading other languages and converting text into audio. Scan marker is also called a digital highlighter    \\
 
Global Positioning System (GPS)  &  GPS generate the real location of the installed objects.  GPS may be installed in school buses. This will automatically generate a message as the bus comes closer to the station. This will save time.     \\
 
\bottomrule    
\end{tabularx}
\end{table*}


\paragraph{\textbf{Smartphones:}}
Research shows that in the coming years, massive smart devices will be connected to the internet. Now, even in developing countries, usually, every family has a smartphone. Flipped classrooms, which are becoming a popular teaching methodology depend on smartphones \cite{SEI149}. 

\paragraph{\textbf{Sensors and actuators:}}
Different types of smart sensors may be installed in the institutions. These sensors monitor and sense the intuition for different parameters in real time.  Actuators are action devices, which are directed by a central control system for any action. 

\paragraph{\textbf{Cameras:}}
Cameras are installed in the institutions and classrooms for the monitoring system; which sends their data in real-time towards the fog layer. Cameras increase the security and transparency of the school.

\paragraph{\textbf{Sound frequency detector:}}
Smart frequency sensors may be installed in the classrooms. This continuously monitors the voice level in the classrooms. When the voice level increases to a certain level, a notification is displayed at the head office.

\paragraph{\textbf{Smart speaker:}} 
The smart speaker provides interactive actions. These act like smart devices and provide a facility of Wi-Fi and Bluetooth for connectivity.  Smart speakers may be used to interact with learners and teachers. To give a special message to any class or to broadcast it towards all classes. Smart speakers may be connected to Alexa and programmed to the institution's timetable for automatic  notification \cite{SEI64}.

\paragraph{\textbf{Voice and gesture activated Virtual assistants:}} 
Controlling the smart environment with the traditional way of device control limits the potential use of smart devices, and it can be more tiring for the user. The voice and gesture-activated smart environment along with \ac{ai} enhance the potential use. For example, Alexa works with spoken instruction, developed by Amazon and uses speech recognition technology to communicate with smart applications and human users. With speech recognition technology, it can automatically help teachers and students. Alexa can be connected to local servers for detailed reports. If necessary, it can also announce the information in speech form \cite{SEI63}.

\paragraph{\textbf{Smart glasses:}}
Among modern inventions, smart glass is one of the portable wearable devices which is capable of handling a wide range of smart activities. Smart glasses are widely considered a next-generation smartphone. Primarily, it is used for human-machine interactions \cite{SEI51,SEI66}.

\paragraph{\textbf{Fingerprint:}}   
A fingerprint is a mark left by the human finger. Attendance systems and crime investigations are important applications. Along with \ac{rfid} and \ac{ai} technology, this technology is popularly used for learners and staff attendance \cite{SEI45, SEI209}.

\paragraph{\textbf{Wearable technology:}}   
Wearable electronic devices, such as smartwatches, smart bands, smart glasses, and smart neckless are used for various purposes. These devices can play a vital role in the smart education system. In smart education, this may be used as a sensor, tracker or timetable notifier, etc.  Smartwatches are probably the most widely available device of wearable technology. It offers almost all the facilities of smartphones. It has the installation of GPS, microphone, camera, and other sensors that can be used to track the child in need. The wearable technology market is expected to grow up to 1.1 billion by 2022 \cite{SEI176}.

\subsection{Fog computing}
\label{fog}
All data coming from \ac{iot} sensors and actuators are stored and processed on servers working on the fog layer. The reason behind using fog computing is to minimize the load on the network. A big stream of data is produced  by smart devices which are used for smart reporting and decision \cite{SEI181}. 
Local servers are placed in the institution and all the \ac{iot} layer devices are attached to this. The local fog devices store the data, which creates a distributed cache network; the fog devices also process the incoming data and if needed, send it to the next cloud server \cite{SEI162}\cite{SEI180}. 

\subsection{Cloud computing}
\label{cloud}
This layer is the cloud of fog servers and other special servers connecting the education system. This connects the entire network to share the institutions’ experiences, official records, videos, etc with each other \cite{SEI140}.  The higher authority may  also use the cloud layer to get detailed information from registered institutions. This layer uses smart algorithms to facilitate top authorities to actively monitor the institutions \cite{SEI61,SEI211}. 

\section{Structure of smart education }
Smart education received very little attention, therefore, in the literature, there is no specific definition, structure, or environment. This section reviews the existing literature related to the concept of smart education. 

\subsection{Concept of smart learning}
 Smart education is a multidisciplinary domain, covering ICT and education.  Smart education is an educational learning and management paradigm, in which smart technologies (e.g. IoT, AI, and 5G) are applied to make it more efficient and engaging. Smart education provides a digital environment to help students, parents, teachers, and administrators to improve learning engagement and motivation \cite {SEI156,SEI163}.

\citet{SEI161} suggested a concept of intelligent learning as follows: first, it is more learner and content-oriented than device-oriented; second, it is efficient, intelligent, and personalized learning based on an advanced IT infrastructure.
Additionally, \citet{SEI164}  utilized cloud computing for smart learning. They believed that smart learning combines the benefits of social learning and learning everywhere, which is a learner-centred and service-oriented educational paradigm, rather than a paradigm focused solely on the use of devices.

\subsection{Smart learning environment}
The smart learning concept provides a self-adaptive, personalized, and self-motivated environment for learners, teachers, parents, and administrators. Smart education is a digital environment enriched with smart devices to enhance learners' motivation and satisfaction. The Journal of Smart Learning Environment was launched in 2014 to facilitate the learners.

\citet{SEI165} clarified that possible criteria for an intelligent learning environment include context awareness, the ability to provide direct and adaptive support to learners, and the ability to adapt the learner's interface and content of the material. This environment facilitates administrators with learning and managing. 
\citet{SEI166} defined a learning environment as a physical environment which facilitates learners' learning. 
Furthermore, \citet{SEI167} state  that the smart learning environment offers the possibility to use the students' free time productively. It plays the role of a coach and guide to knowledge. It facilitates learners by offering self-learning, motivated and personalized services.
\citet{SEI168}  worked on effectiveness, efficiency, engagement, flexibility,  and defectiveness  in the smart learning environment.  

The International Association of Smart Learning Environments is an association that works on smart learning environments to improve learning. They present the smart learning environment in 6 strategies, which follow; (a) infrastructure,  (b) learning tools, (c) learning resources, (d) teaching and learning methods, (e) services for teachers and students, and (f) collaboration between government, businesses and schools \cite{SEI169}.

\subsection{Smart Learning Framework}
In the literature, no specific framework for smart learning is defined, however, some of its subdomains are discussed. \citet{SEI170} proposed a framework (Flipped learning wheel (FLW))  for the flipped classroom. The FLW randomly groups the learner for a particular topic. Learners discuss these topics with each other after that expert groups are created for discussion and at the end, they are assessed.  Authors in \citet{SEI165}, presented a framework for the smart learning environment.  Authors in  \citet{SEI156} presented a framework for smart learning. This framework has three components; smart environments, smart pedagogy, and smart learner.


\begin{figure*}[h]
\begin{subfigure}{0.45\textwidth}
\includegraphics[width=1\linewidth]{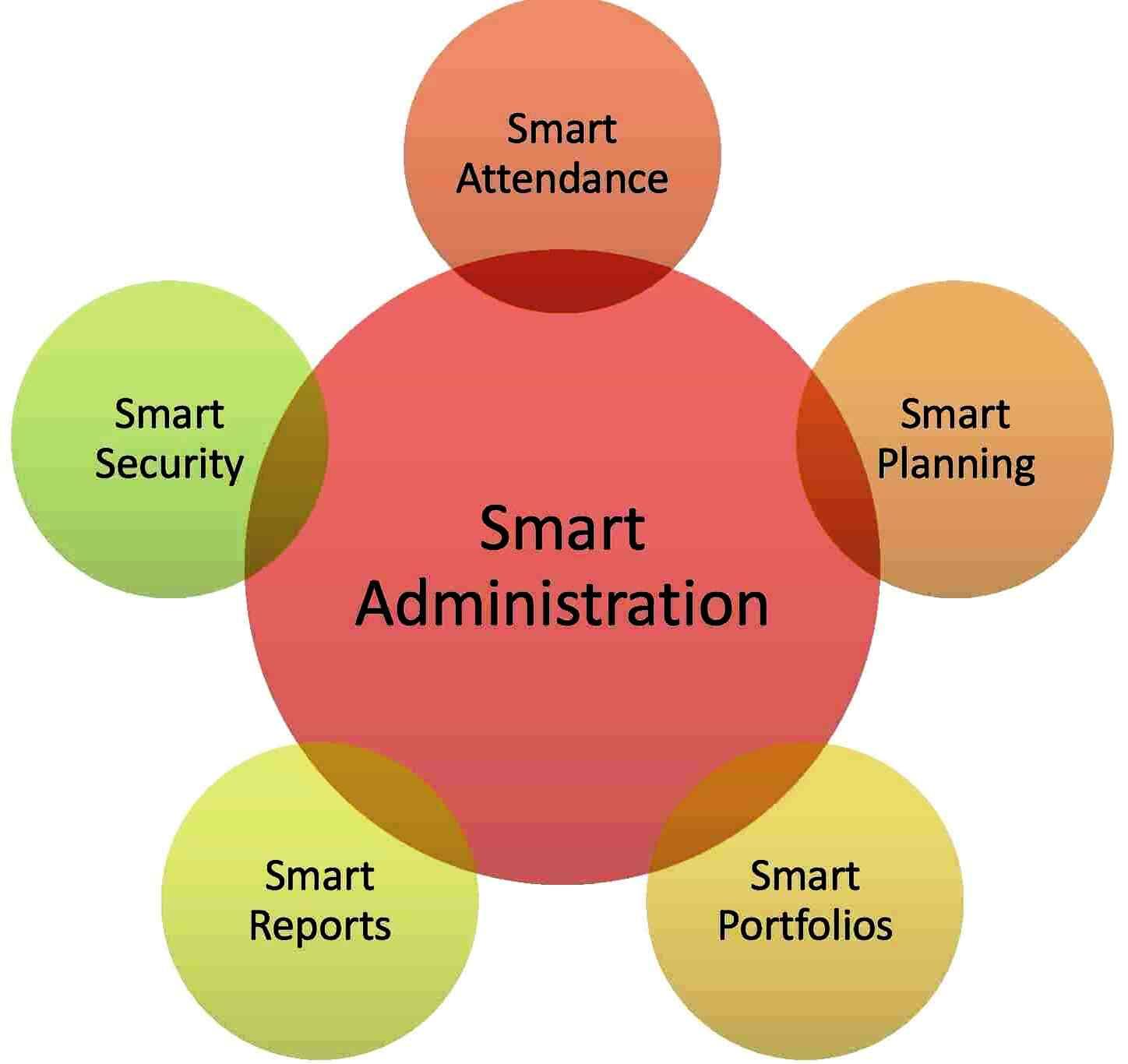} 
\caption{Structure of Smart Administration}
\label{fig:smartadministration}
\end{subfigure}
\begin{subfigure}{0.45\textwidth}
\includegraphics[width=1\linewidth]{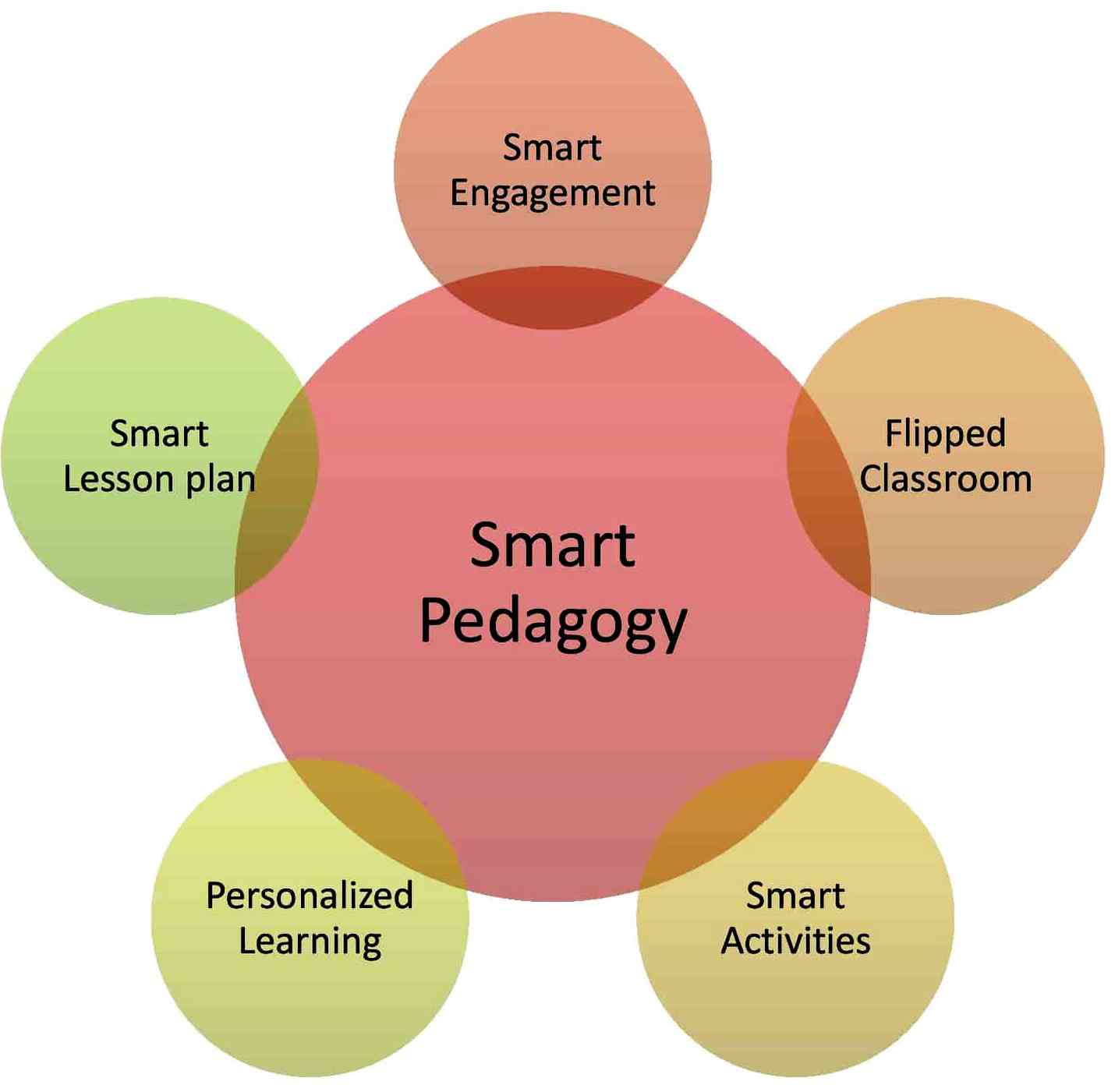}
\caption{Structure of Smart Pedagogy}
\label{fig:smartpedagogy}
\end{subfigure}
\caption{Structure of Smart Administration and Smart Pedagogy}
\label{fig:smartstructure1}
\end{figure*}

\section{Possible uses of  IoT in educational institutions}
\label{app}
Smart devices improve educational institutions in a number of ways. This section explores  in detail, how smart devices help to improve the learning and teaching process.  Fig. \ref{fig:smartaaministration} shows the structure of smart educational institutions and \ref{fig:smartstructure1},  \ref{fig:smartstructure2}  shows the smart solutions to the traditional education challenges.  Tables \ref{tab:npf1} and \ref{tab:npf2} discuss the  possible usage of  smart devices in the education system.

\begin{table*}
\centering
\caption{Smart solution using \ac{iot} and smart system}
 \label{tab:solution} 
 \newcolumntype{b}{X}
\newcolumntype{s}{>{\hsize=.4\hsize}X}
\newcolumntype{c}{>{\hsize=.9\hsize}X}
\setlength{\extrarowheight}{2pt}%
\begin{tabularx}{\textwidth}{@{}s c s@{}} 
\toprule
\textbf{Proposed Solution} &  \textbf{Description} & \textbf{Devices / Technologies needed} \\
\midrule

Smart attendance & \ac{rfid} smart   cards  are  used  to  automate  the attendance  system  of  the  school.  This  mark  the  learner  and teacher's  attendance  as  they  enter  the  class.       &   \ac{rfid}, Cameras, Biometric, \ac{ai}\\

Smart portfolios & learners' assessments, attendance and character record are stored in the database. Parents are notified through applications and messages.   & Server, mobiles phones, \ac{rfid}, Cameras, Biometric, \ac{ai}   \\

Smart pedagogy  &  Different smart methodologies  and devices (e.g, flipped classrooms, video recording, \ac{ar}, and \ac{vr}) are used to creatively and actively engaged the students.  & Cameras, \ac{ar}, \ac{vr}, IR, and smart boards etc   \\

Smart lesson planning  &  To overcome the class issue and to have a look that the teacher is fulfilling lesson planning SOPs; the lesson plan is updated online.   &  Servers, Mobiles phones, PDAs, Laptops  \\

Learner engagement & Engaging the learning by using \ac{iot}, \ac{ai}, AR, VR and other smart systems.    &  Animation, \ac{ai}, \ac{ar}, \ac{vr} and smart glasses  \\

Flipped classroom & In the flipped classrooms, students watch videos at home and do activities at school under teacher observation.      &  Cameras, \ac{ar}, \ac{vr}, Animation   \\

Massive Online Open Courses &  In \ac{mooc}, online courses are provided by Universities and different institutions for students.   & Cameras, Mobile phones, PDAs, Laptops    \\

Smart assessment & Smart testing services are used to take the examination of the learners and then store these results on fog servers.      &  Mobile phones, PDAs, laptops  \\

Smart collaboration &  Video conferencing, \ac{ar}, \ac{vr} and \ac{ai} is playing an excellent role to enhance remote smart collaboration among students.   &  Mobile phones, PDAs, laptops, \ac{ar}, \ac{vr}, \ac{ai}    \\

Personalized learning & In personalized learning, courses and topics are assigned to the students as per their activities analysis.     &  Fog and cloud servers, \ac{ai}, Data mining   \\

Smart administration & Smart devices and reports are used by the administrator to smartly manage the employees, students and other related activities.    &  Fog and cloud servers, \ac{ai}, Data mining   \\

Learner Analysis & Learner analysis used the smart devices, their activities and learning reports to smartly analyse the learner and generate their report.     &  Fog and cloud servers, \ac{ai}, Data mining   \\

\bottomrule    
\end{tabularx}
\end{table*}

\subsection{Smart administration}
Monitoring teachers' and learners' progress in institutions is  critical. An important part of operating a school is keeping track of how well the teachers are doing and evaluating how effective are they in their work. With the proper evaluation of teachers and institutions, the performance can be improved to a significant extent as well. The smart system can transparently manage all these issues. Furthermore, administrative tasks have increased the burden on teachers. This burden may be minimized by smart systems, so teachers may concentrate on teaching only \cite{SEI106}. Fig. \ref{fig:smartaaministration}  shows the structure of smart administration.

\paragraph{\textbf{Smart attendance:}}
Manual or fingerprint attendance is time-consuming. It is also much more boring to mark attendance before every class. At the school level, it is even not possible where there are eight periods of 40 minutes. Where each class contains a minimum of 60 learners. \emph{\ac{rfid} smart cards} may be used to automate the attendance system of the school. This mark the learner and teacher's attendance as they enter the class. The smart system  notifies the teacher and principal how many learners are present in the class. Smart cards  also make sure of the teachers' presence and timing in the class. Technology is invisible and works silently. 
Smart attendance reports the parents on their arrival or departure from school, keeping them aware of their child. This  also upgrades the learner log when he/she enters or leaves the library, cafe or labs, etc \citet{SEI200, SEI210}.

\paragraph{\textbf{Smart portfolio:}}
learners' assessment, attendance, and character records are stored in the database. On new complaints, parents are notified through applications and messages. Teachers or parents may check throughout the summary of  learners. This  helps parents and teachers to diagnose the issues with any learner. The administrator of the institution monitors the institution in real-time. They can check any teachers’ lesson plans, homework, courses covered and results, etc.  The homework and lesson plan may also be directly viewed by parents to introduce the flipped classroom environment. Video as a sensor may be used for classroom  close observation \cite{SEI80}.

\paragraph{\textbf{Smart reports:}} 
The administration takes decisions on the reports forwarded to their offices. With manual reports, there are major chances of report manipulation, which leads to wrong decisions and actions. The introduction of the smart system minimizes corruption and illegal report manipulation. 

\paragraph{\textbf{Smart security:}}
Thousands of students and staff study and work in educational institutions. With limited security equipment, they are  easy targets for terrorists. The intelligent system plays a major role in security by installing cameras and other sensors. The rise of \ac{ai} further improves the performance of this equipment \cite{SEI72,SEI201,SEI202}.

\subsection{Smart pedagogy}
Pedagogy is a set of skills used for effective teaching in the classroom. It is not only limited to teaching but also covers the psychology skills to read and understand learners \cite{SEI107, SEI199}.  Figure \ref{fig:smartpedagogy}  shows the structure of smart pedagogy.

\paragraph{\textbf{Smart  engagement:}}
Learners’ motivation and engagement is the most appealing element in the teaching process. Learner engagement increases the critical thinking of the learners. In traditional classes, it is observed that minimum learners are engaged. To maximize the quality of education, learners' engagement must be enhanced.  Authors and teachers have already been using \ac{iot}  to keep the learners motivated and engaged in some particular subjects \cite{SEI46, SEI87}.

\paragraph{\textbf{Flipped classrooms:}} 
The fog layer helps in flipped classrooms. \emph{Flipped classrooms} are a new emerging teaching methodology in which teachers record videos for learners. They watch these videos at  home and do related activities in classrooms under the teacher's supervision. The fog server stores lecture videos and homework and also broadcast them to parents’ smartphones. Flipped classrooms may be applied using Bring Your Own Device (BYOD) agenda \cite{SEI150, SEI198}. 

\paragraph{\textbf{Personalized learning:}}
One of the biggest issues with traditional education is that every learner has to cover the same course content. There is no age for slow learning. Information and Communication Technology (ICT) gives a facility of personalised courses for every learner. They can learn in their own space. This increases the interest and engagement of the learners.  

\paragraph{\textbf{Smart lesson planning:}}
 In public teachers, it is usually assumed that "you will force me to enter into the class, but you cannot force me to teach there", therefore, to overcome this issue and to have a look that the teacher is fulfilling lesson planning SOPs; the lesson plan should be created smartly.  Principals, learners, and parents may review the lesson plan of every teacher and class. In case of low performance or violation of lesson planning, the system will automatically highlight the mishaps. Smart and transparent lesson planning is very important because it  makes the teachers' work transparent to principals, parents, and learners.  Previous lesson plans or borrowed lesson plans may be used which will save time.

\paragraph{\textbf{Smart activities:}}
Learning by doing increases the retention and understanding of the learners. Teaching other learners in activities increases retention by up to 90 percent. Smart activities (e.g, smart applications, games, and smart objects) keep the learners engaged with learning. China installed 400,000 3D printers across elementary schools to efficiently teach the concepts \cite{SEI124}.

\subsection{Smart assessment} 
\begin{figure*}[h]
  \centering
\begin{subfigure}{.45\textwidth}
  \includegraphics[width=1\linewidth]{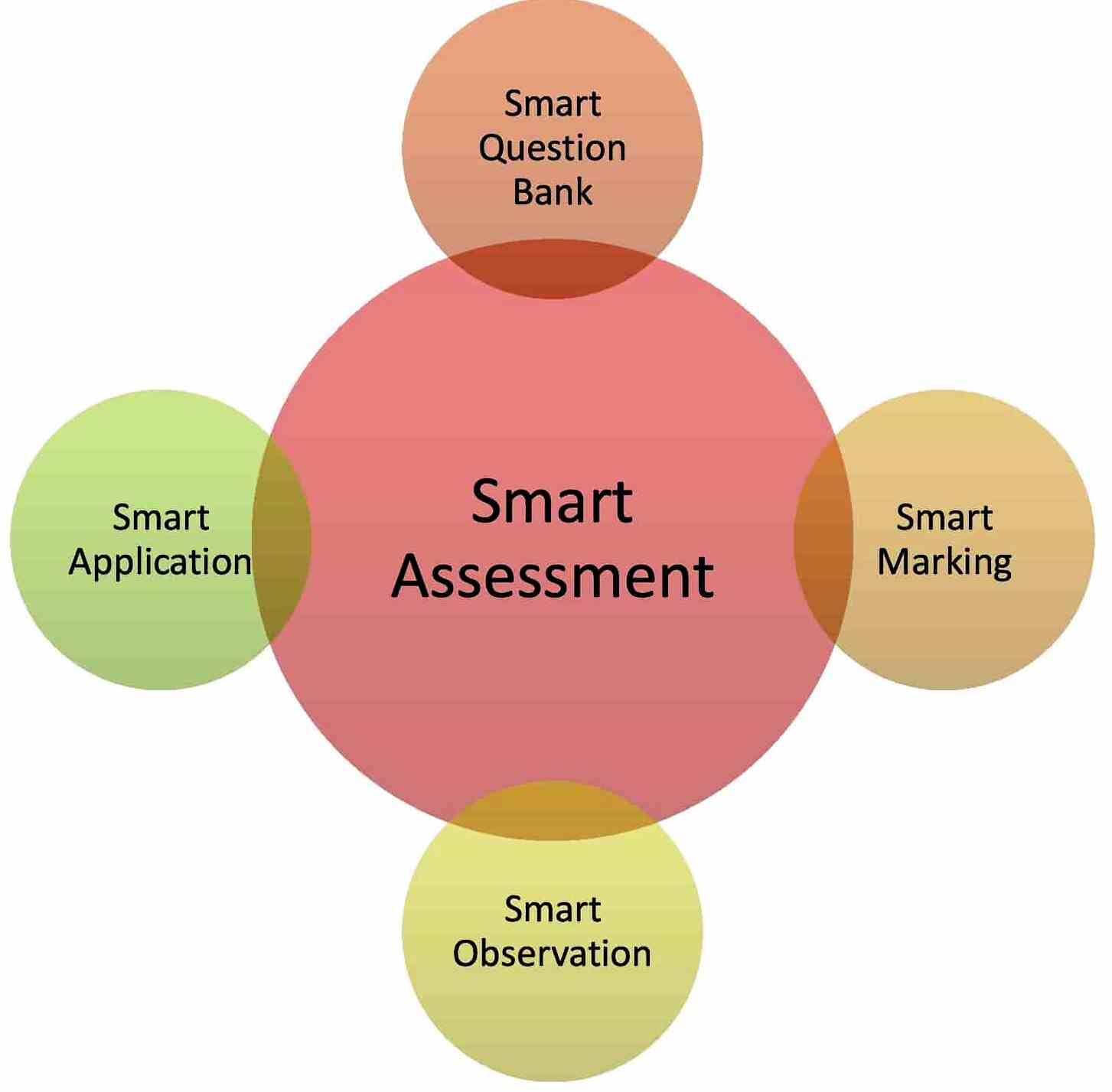}
\caption{Structure of Smart Assessment}
\label{fig:smartassessment}
  \label{fig:rdc1}
\end{subfigure}
  \centering
\begin{subfigure}{.45\textwidth}
  \includegraphics[width=1\linewidth]{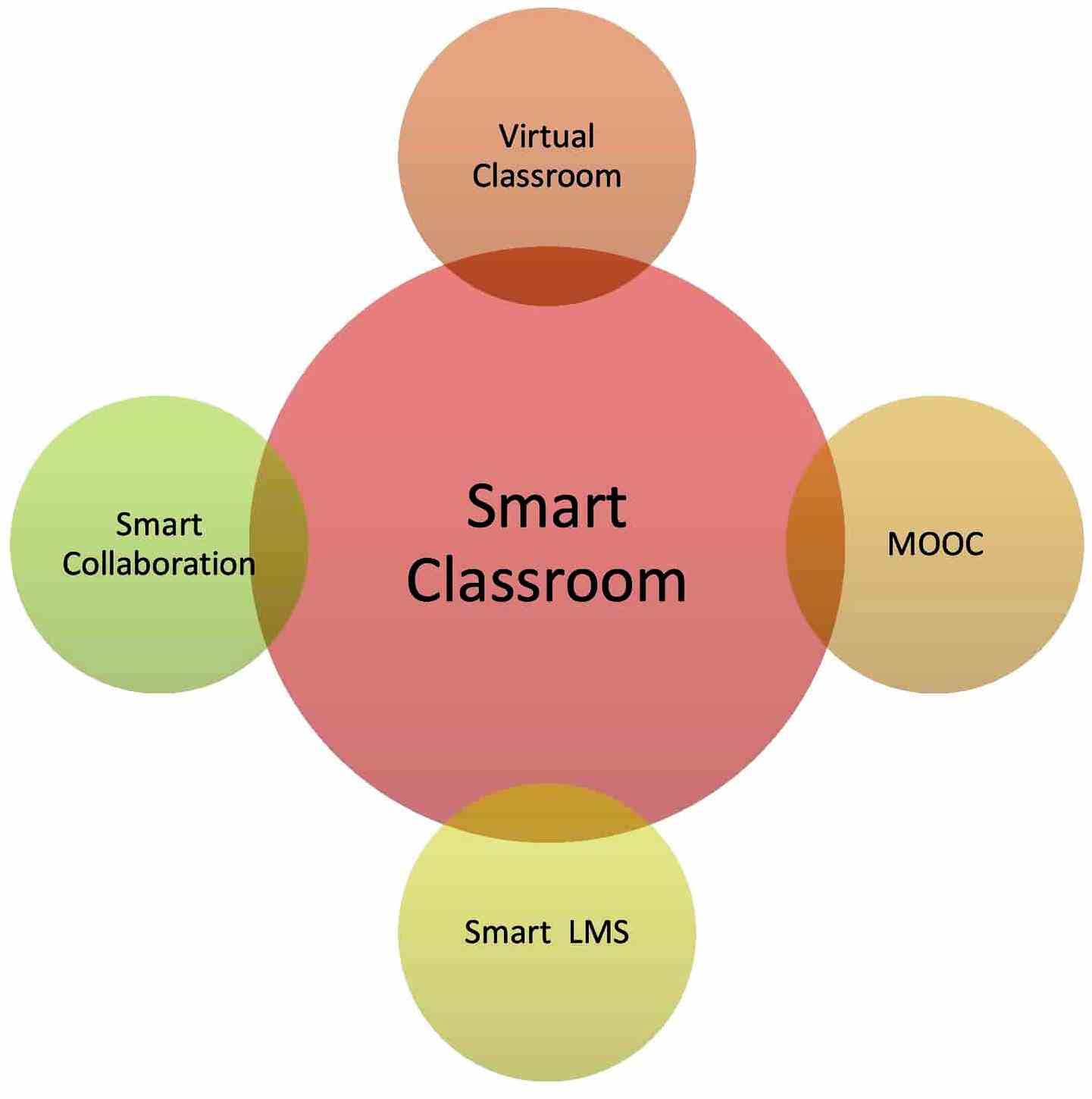}
\caption{Structure of Smart Classroom}
\label{fig:smartclassroom}
\end{subfigure}
\caption{Structure of Smart Assessment and Smart Classroom}
\label{fig:smartstructure2}
\end{figure*}

\emph{Assessment transparency} is  important to evaluate teachers and learners. Smart testing services are used to take the examination of the learners and then to store these results on fog servers. These records help in transparent reporting and decision. Fig. \ref{fig:smartassessment}  shows the structure of smart assessment.

\paragraph{\textbf{Smart questions bank:}}
A database is created for each topic for the smart assessment application. Teachers add questions to the database for each exam and automatically prepare the paper by choosing specific questions. The quality of these questions is monitored and improved over time. These smart question banks save teachers time in creating and marking tests.

\paragraph{\textbf{Smart marking:}}
During the creation of the smart question banks, their answers are also uploaded to the database. Therefore, during checking (this can be a  code or an online exam), the application automatically checks the paper for correct answers. This saves time and effort. It also minimizes the exam cost. 

\paragraph{\textbf{Smart observations:}}
Instead of taking a written or oral test, learners are passively observed for their behaviour and tendencies. Observation is an important part of the assessment and important decisions are made on its behalf. Psychological analysis is a crutial factor in teaching. Teachers are trained in their training to read their learners and to behave according to their mental state. The intelligent system and \ac{ai} technologies can be used to read and engage learners. This will help teachers to deal with troublesome learners. It can also be used to group learners accordingly \cite{SEI47}.

\begin{table*}[h]
\begin{center}
\caption{Criteria for studies assessment}
\label{tab:criteria}
\newcolumntype{a}{>{\hsize=.1\hsize}X}
\newcolumntype{b}{>{\hsize=.3\hsize}X}
\newcolumntype{c}{>{\hsize=.8\hsize}X}
\setlength{\extrarowheight}{5pt}%
\begin{tabularx}{\textwidth}{@{}a  b   c@{}}
\toprule
\textbf{Symbol} &  \textbf{Criteria } & \textbf{Criteria Definition} \\
\midrule

C1  &   Review of related studies    &  This criteria checks whether the authors compare this study with the related studies and projects  or not.        \\
C2  &   Challenges in traditional education  &    Criteria two evaluates that the authors consider the  challenges that exist in the traditional system.    \\
C3  &   Proposed possible solutions    &   Checks that authors proposed possible solutions to the challenges that exist in the traditional education system.    \\
C4  &   Smart human resources    & deals that how human resources (learners, parents, teachers and administrators)  work with the smart system.      \\
C5  &   Possible resistance     &   Every new invention has resistance in the community. What is the possible resistance to smart education is covered in criteria 5.     \\

\bottomrule
\end{tabularx}
\end{center}
\end{table*}

\subsection{Smart classroom}
Smart classrooms use smart technologies to help with teaching and learning. It provides interactive whiteboards for smart interaction, smart audio and video ads, uses smart and comfortable chairs, \ac{vr} and \ac{ar} to teach the 3d concepts etc.  Rather than manual adjustment, it automatically adjusts the classroom lighting and temperature, and the smart systems automatically manage it. Smart cameras also assist in reading and behavioural assessment.   Fig. \ref{fig:smartclassroom}  shows the structure of the smart classroom

\paragraph{\textbf{Smart collaboration:}}
\emph{Video conferencing} is a need of today's educational institutions. This is used to connect different intuitions with each other. Learners and teachers interact with each other for idea sharing. \ac{iot}, \ac{ai}, and \ac{vr} may bring a dramatic improvement in collaboration. Smart glasses and \ac{vr} may be used to evaluate the training sessions. 

\paragraph{\textbf{\ac{mooc}:}} 
\emph{\ac{mooc}} courses are offered by a number of world top Universities and millions of learners are taking part in these courses. Smart devices (e.g., virtual reality sets, smart glasses and Alexia etc) may be utilized to enhance this learning (in terms of smart teaching and smart  assessment).  

\paragraph{\textbf{Virtual classrooms:}}
Instead of attending classes physically, smart LMS is used to attend video lessons virtually from home. It is a technique that saves time and money. The rise of \ac{vr} and \ac{ar} reality improves the quality of virtual lessons by teaching 3D concepts \cite{SEI146}.

\paragraph{\textbf{Cyber-physical smart systems:}}
\ac{cps} encompass a variety of technologies; including IoT, cloud computing, fog computing, and swarm intelligence \cite{SEI172, SEI175}. \ac{cps} is a robotic system that works automatically to perform any task. The merged reality or augmented reality can be used for conducting experiments and other practical work, where physical prance is required. With the use of augmented reality, students would be able to conduct experiments by controlling the robotic setup from home, similar to solutions already developed for performing surgery from distance. However, adopting such sophisticated solutions at different levels of education institutions will require a modified or customized setup \cite{SEI119}. 

\paragraph{\textbf{Smart monitoring and tracking:}}
Smart technology facilitates parents, teachers, and administrators to smartly track learners. The \ac{ai}  read the faces for expressions (e.g, sadness, happiness, surprise, fear, anger, and disgust) of the learners and describes their emotions. This will  help the teachers to identify these students who feel boredom in the classroom. On the basis of this analysis, teachers may take decisions for warmer or energizer etc \cite{SEI177}.

\paragraph{\textbf{Smart Video Splitting:}}
Virtual classrooms, conferences, and MOOC courses have intensified the streaming use in teaching and learning. Leading platforms, e.g.,  Google Meet \cite{SEI142}, and Zoom \cite{SEI141}, etc are used to connect with teachers and learners. The classes or conferences continue long and the hours-long videos are recorded. In these long videos, it is hard to search for the required content nor these contents are easily conveyed. 
Various smart video techniques (e.g., video skimming \cite{SEI189}, summarization \cite{SEI190}, and condensation \cite{SEI191}) are used to split the videos based on specific content.  
This \ ac {ai} based algorithms may be used to separately break the video as per the gestures or speech recognition and save it as a separate video. This will make it the learners and other users easy to search the required content easily \cite{SEI188}.

\begin{table*}[h]
\begin{center}
\caption{Comparative analysis of related studies}
\label{tab:summary}
\newcolumntype{b}{X}
\newcolumntype{s}{>{\hsize=0.3\hsize}X}
\newcolumntype{c}{>{\hsize=0.08\hsize}X}
\newcolumntype{d}{>{\hsize=0.8\hsize}X}
\setlength{\extrarowheight}{5pt}%
\begin{tabularx}{\textwidth}{@{}s c   c  c  c  c  c  c @{}}
\toprule
\textbf{Paper} &  \textbf{Year }   &  \textbf{C1 } & \textbf{C2} & \textbf{C3} &  \textbf{C4 } & \textbf{C5}  \\
\midrule
	
  \citet{SEI78} & 2018 &	{\cmark}	&	{\xmark} 	&	{\cmark}	&	{\xmark}	&	{\xmark}	  \\
 \citet{SEI79} & 2018 &	{\cmark}	&	{\xmark} 	&	{\cmark}	&	{\xmark}	&	{\xmark}		\\	
 \citet{SEI73} & 2019 & 	{\cmark}	&	{\xmark} 	&	{\cmark}	&	{\cmark}	&	{\cmark}		\\
 \citet{SEI74} & 2019 &	{\cmark}	&	{\xmark} 	&	{\cmark}	&	{\xmark}	&	{\xmark}		\\
 \citet{SEI75} & 2019 &	{\cmark}	&	{\xmark} 	&	{\xmark}	&	{\cmark}	&	{\xmark}		\\
 \citet{SEI76} & 2019 &	{\cmark}	&	{\xmark} 	&	{\xmark}	&	{\xmark}	&	{\xmark}		\\
 \citet{SEI77} & 2019 &	{\cmark}	&	{\xmark} 	&	{\cmark}	&	{\xmark}	&	{\xmark}		\\
 
Afzal et al & 2020 &  {\cmark}  & {\cmark} & {\cmark} & {\cmark} & {\cmark}  \\

\bottomrule
\end{tabularx}
\end{center}
\end{table*}

\section{Challenges  and future directions }
\label{challanges}
Rather than the tremendous benefits of a smart system, there are several challenges and resistance to its implementation. These challenges and resistances are classified into two categories; (i) computational challenges and (ii) social challenges and resistance.

\subsection{Computational challenges}
Computational challenges run parallel to the installation and integration of smart systems. The general challenges commonly encountered with smart devices are;  network, battery,  or communication failures. The next streong challenge is that the built-in functions stop working and these minor issues require experts. Then comes the code updating. If changes to the program need to be updated, the user cannot do this easily \cite{SEI96,SEI194}. 
Some specific challenges related to smart education are as follows: 

\paragraph{\textbf{Internet connectivity:}}
Faraway areas are still deprived of internet connection. This system can not be introduced where there is no internet connectivity or slow internet access. Moreover, smart systems work with real-time input, therefore, in case of a slow internet connection, this may not be installed \cite{SEI84}. 
In case of slow internet, fog servers can help \cite{SEI133}.  The data may be initially stored on fog computing and can be uploaded to the cloud at a slow speed, similarly, the downloads may be continued. 

\paragraph{\textbf{Privacy issues:}} 
User privacy is another important issue. It is a known fact that smart services require data to enhance personalized experiences. Personal data sharing with service providers and manufacturers can lead to a privacy breach. There are chances that the institutions' data and camera stream may be hacked by someone else. This data may be used for negative purposes \cite{SEI126}. Protecting the data from illegitimate access requires authenticated data access \cite{SEI126a, SEI126b}.   
Privacy has been widely researched and there are a number of solutions discussed in the literature which can be employed in smart education \cite{SEI134,SEI134b,SEI135}.  

\paragraph{\textbf{Compatibility and Interoperability:}}
Devices compatibility and interoperability rise in every integration, especially at the school level to handle such technical issues. This is the third most powerful challenge, which resists the implementation of smart education \cite{SEI81}. 
The education sector is a big market. Everyone is connected (one way or other) to the education system. EduTech can be expedited to develop a smart education system commercially \cite{SEI136}.   This will end the compatibility and intractability issues. 

\paragraph{\textbf{Data pollution:}}
These massive devices produce massive data and most of the data is not in use. The learners detract from such large internet data. Particularly, the learners are addicted to social media and they spend hours consuming this data pollution \cite{SEI81}. Therefore, the challenge here is how to avoid learners from data pollution \cite{SEI123}.
  To overcome this challenge, a learners' monitoring module may be added to the smart education system to monitor the view history of the learners \cite{SEI137}. This history should be transparent to learners, parents, teachers, and administrators. With this log, learners will limit their activities to learning material only. 
  
  \paragraph{\textbf{Artificial Intelligence:}}
The major issue with \ac{ai} is the lack of trust because it is not so mature to correctly read the environment or to hire high-level experts. Therefore, this may input the wrong data or anyone can give the wrong input. For example, students may give sad expressions to get warmer or break.  Furthermore, the \ac{ai} will increase the cost, which would not be paid from the budget \cite{SEI178}.
Because of the challenges mentioned above, the use of \ac{ai} should be limited based on budget and accuracy.

\subsection{Social challenges}
Despite the countless benefits, the smart system faces many challenges from the social side. The most general one is the use of modern technologies because the new person does not have enough information to use these systems \cite{SEI125}. The specific issues related to social challenges are as follows:

\paragraph{\textbf{Employees resistance:}}
Every new technology implementation has resistance from people, especially from employees. Monitoring is unfriendly for human beings and teachers do not like such a system where they are monitored on a real-time basis. Therefore, the major challenge towards smart institutions is the employees of educational institutions \cite{SEI82, SEI184}. 
Special incentives and motivation are needed to overcome the employees' resistance. One of the reasons for employee challenges is passive communication between employees and administrators. Therefore, the system may be further extended to include the proper correspondence system in the smart education system \cite{SEI25}. This will not only minimize the employees' resistance but also  make them happy for handling their issues efficiently.

\paragraph{\textbf{Use of smart systems:}}
In most countries, teachers are not well aware of smart devices and smartphones. They are not able to use these devices for assessment and other tasks. They will create hurdles in successfully implementing this system.
This challenge can be handled through training, however, disinterest may lead to issues. Therefore, a post-training assessment module can be added to the smart system to monitor the understanding of the employees. In case of a lower score, the training me to be rescheduled \cite{SEI26}.

\paragraph{\textbf{Lack of funds:}}
In underdeveloped countries, lack of funds is the biggest problem. Institutions usually do not even have  funds to build classrooms for learners or buy furniture. In such areas, where schools are allocated, a very low budget will face obstacles towards smart institutions \cite{SEI85}. Especially, the \ac{cps} system needs massive funds for installation and public institutions would not be able to support it. 
However, education is a big market and everyone is connected to it directly or indirectly.  The system production on a large scale can reduce the cost \cite{SEI127}.

\paragraph{\textbf{Cultural issues:}}
In some cultures,  girls are not liked to capture photos or videos. In such like situations, people may resist installing cameras or sensors in educational institutions. Furthermore, some parents do not allow their children to use smart devices. Flipped classroom ideas may not work in such situations. 
However, the smart system also helps to overcome these issues by protecting the culture. For example, female students can attend the virtual classroom without facing male teachers \cite{SEI50,SEI128}.    

\paragraph{\textbf{Technology addiction:}}
The research article \cite{SEI111}, concluding different studies shows that kids are heavily addicted to mobile phones. Therefore, one of the challenges with digitization is that students may addict to different applications, instead of giving time to learning.  
Overcoming and controlling technology addiction is a primary concern because this negatively affects the learners instead of facilitating their learning. The learner monitoring module can be embedded with the smart system to monitor the learners and check their watch history. This way, learners will try to control their watch history \cite{SEI129}.

\paragraph{\textbf{Digitizing books:}}
All this will need to digitize the syllabus and make animation videos for AR and VR. This will need massive cost, time, and expertise \cite{SEI206}. 
Creating animated stuff is complicated, however, an intelligent system can be created for teachers (not experts in technology) to create animated videos for learners with very little effort \cite{SEI130}. This can be embedded with the teachers' application to manage the stuff along with their lesson plans, etc.    

\paragraph{\textbf{Training:}}
Teachers will need to be trained in the smart system. This will also need massive cost and time. It is also possible that teachers may not get the required expertise after training and they may get failed to use the smart system \cite{SEI207}. 
For the easiness of the teachers and administrators, an automated module can be created and embedded with the smart education system to train the teachers \cite{SEI131}. The teachers can give time as per their schedule to complete the training and go through assessments to get the certificate.

\section{Comparative analysis}
\label{ca}
This section compares this study to related studies. Unfortunately, very limited review papers are available on \ac{iot} in education. We used the criteria given in Table \ref{tab:criteria}, to compare this study with related studies. C1 checks whether or not the studies and associated projects are covered; C2 discusses the challenges in the traditional educational system; C3 covers the possible solutions for challenges in traditional education systems, C4 addresses the working of human resources in the smart education, and finally, C5 evaluates the resistance in implementing the smart institution framework. 
 
\citet{SEI78} presented a systematic review of information system implementation in the higher education system.  They reviewed the related studies and explored the challenges in existing information management systems.
\citet{SEI79} reviewed the role of \ac{iot} in the educational system. They further proposed some solutions to the existing challenges using of \ac{iot}.
\citet{SEI73} reviewed the literature related to smart universities. They also presented the use of \ac{iot} in making campuses smart. 
\citet{SEI74} explored the opportunities and challenges of using \ac{iot} in education systems. They explored the literature and presented some solutions towards smart education. 
 \citet{SEI75} reviewed related studies and partially discussed the use of blockchain technologies in university campuses to ensure accountability, transparency, and  cyber-security. 
 \citet{SEI76}  reviewed and discussed the main seven categories of \ac{iot} in education, including  Intelligent Tutoring System (ITS), smart campus, Big Data in Education (BDE), knowledge graph, educational robots, virtual teachers, and personalized education.
 \citet{SEI77} reviewed the literature about the use of \ac{iot} in \ac{mooc} and possible challenges towards integrating  of \ac{iot} in \ac{mooc}. 

Comparatively to these surveys and reviews, the proposed study is very wide and deep. This covers the literature including research studies and different projects; it explores the real  issues faced by learners, teachers, and administrators in educational institutions. On the basis of these data and the specification of \ac{iot} and other smart technologies, solutions are proposed for the issues in traditional systems. It also explores the smart human resources model which explains how human resources will work  with the smart educational framework.  Every new technology faces resistance from the community and employees, therefore, this study explores the possible resistance to the implementation of smart education. 

Keeping in view the above contributions and existing literature, it is crystal clear that this study is the first novel investigation which deeply investigates the integration of \ac{iot}, 5G, and \ac{ai} in educational systems.

\section{Conclusion }
\label{conclusion}
This article explored the \ac{iot} and smart devices integration in education and proposed a possible framework for  smart education.  This further  explored  the issues in educational institutions  (running in the traditional way) and the possible solution to cover these challenges with smart devices. It has been observed (teacher training and mentoring session by the author) that students are not properly engaged, nor the standards are properly followed. Moreover, the teachers' time is wasted on administration tasks, which may be managed smartly. Smart devices convert the traditional activities of the institution to smart activities such as smart attendance, smart reporting, smart pedagogy, smart lesson planning, smart learner engagement, smart reporting, flipped classrooms, smart assessment, and smart security. This article concluded the related research articles and investigated the issues in the traditional education system and  proposed the possible solutions using \ac{iot} and \ac{ai}.


\bibliographystyle{ACM-Reference-Format}
\bibliography{references}


\end{document}